\documentclass[11pt]{article}
\usepackage[affil-it]{authblk}
\usepackage{amsfonts}
\usepackage{amsmath}
\usepackage{hyperref}
\usepackage{fullpage}
\usepackage{stackrel}
\usepackage{amssymb}
\usepackage{graphicx}
\usepackage{tikz}
\usetikzlibrary{matrix,chains,positioning,decorations.pathreplacing,arrows}
\usepackage{qcircuit}
\usepackage{float}
\usepackage{braket}
\usepackage{subcaption}
\usepackage{mathrsfs}

\newcommand\DD{\mathop{}\mathrm{D}}
\newcommand\ii{\mathop{}\mathrm{i}}
\newcommand\ee{\mathop{}\mathrm{e}}
\newtheorem{rem}{Remark}
\newcommand{\GG}[1]{{\color{green!50!black} #1}}

\newcommand{\R}{\mathbb{R}}

\newcommand\dd{\mathop{}\mathrm{d}}

\providecommand{\xv}{\boldsymbol{x}}
\providecommand{\rme}{\mathrm{e}}
\providecommand{\rmi}{\mathrm{i}}
\providecommand{\CX}{\mathrm{CX}}
\providecommand{\wv}{\boldsymbol{w}}
\providecommand{\de}{\mathrm{d}}
\providecommand{\Min}{m_\mathrm{in}}
\providecommand{\Mout}{m_\mathrm{out}}
\providecommand{\Hi}{\mathscr{H}}

\begin{document}

\title{On the capacity of a quantum perceptron for storing biased patterns}

\author[1,2]{Fabio Benatti}
\author[3,4]{Giovanni Gramegna}
\author[5,6]{Stefano Mancini}
\author[7,8]{Gibbs Nwemadji}

\affil[1]{Dipartimento di Fisica, Universit\'{a} di Trieste, Strada Costiera 11, I-34151, Trieste, Italy}
\affil[2]{Istituto Nazionale di Fisica Nucleare, Sezione di Trieste, Strada Costiera 11, I-34151, Trieste, Italy}
\affil[3]{Institute for Theoretical Physics, Universitat T\"ubingen, Geschwister-Scholl-Platz
72074 T\"ubingen, Germany}
\affil[4]{Dipartimento di Fisica, Università di Bari, I-70126 Bari, Italy}
\affil[5]{Scuola di Scienze e Tecnologie, Universit\`{a} di Camerino, I-62032 Camerino, Italy}
\affil[6]{Istituto Nazionale di Fisica Nucleare, Sezione di Perugia, Via A. Pascoli, I-06123 Perugia, Italy}
\affil[7]{Theoretical and Scientific Data Science Group, Scuola Internazionale Superiore di Studi Avanzati (SISSA), Via Bonomea 265, I-34136 Trieste Italy}
\affil[8]{Quantitative Life Sciences, The Abdus Salam International Centre for Theoretical Physics (ICTP), 34151 Trieste, Italy}

\date{\today}

\maketitle

\begin{abstract}
Although different architectures of quantum perceptrons have been recently put forward, 
the capabilities of such quantum devices versus their classical counterparts remain debated. 
Here, we {consider random patterns and targets independently distributed with biased probabilities and}
investigate the storage capacity 
of a {continuous} quantum perceptron model that admits a classical limit, thus facilitating the comparison of {performances}. {Such a more general context extends a previous study of the quantum storage capacity where using} statistical mechanics techniques in the limit of a large number of inputs, it was proved that no quantum advantages are to be expected concerning the storage properties. This outcome is due to the fuzziness inevitably introduced by the intrinsic stochasticity of quantum devices. We strengthen such an indication by showing that the possibility of indefinitely enhancing the storage capacity for highly correlated patterns, as it occurs in a classical setting, is instead prevented at the quantum level.
\end{abstract}


\section{Introduction}

Machine learning aims at building methods that are able to make predictions or decisions based on sample data, without being explicitly programmed to do so. Quantum information theory studies the storage and transmission of information encoded in quantum states. Nowadays these two disciplines are becoming intertwined giving rise to the field of quantum machine learning. 

The flow of ideas runs both ways: on the one hand applications of machine learning techniques are envisaged to analyze quantum systems \cite{Cetal19,Ketal21}, on the other hand, the implementation of machine learning concepts on quantum hardware is also actively investigated \cite{SSP14,DB18,Machiavello19}.
Along this latter avenue quantum advantages are expected in terms of higher storage capabilities and {an increased information processing power}~\cite{W14,Letal21,Hetal21,Ban21}.

The task of precisely comparing the power of quantum and classical neural networks as probabilistic models for information processing and storage is thus becoming pressing. In particular, the issue of determining precisely the storage capacity of the most elementary constituent of a neural network, namely the perceptron~\cite{Rosenblatt57}, has been addressed in the classical scenario without referring to any specific learning rule using several approaches, ranging from combinatorics~\cite{Cover,Venkatesh92} to statistical  mechanics methods~\cite{G88,Shcherbina2003,Tal,Engel}.
The latter has been used recently to generalize the calculation to some models of quantum perceptrons~\cite{Artiaco21,GKL21,BGM22}. However, the results depend on the specific model used (see e.g.~\cite{GKL21} and~\cite{BGM22}, based respectively on the models~\cite{Machiavello19} and~\cite{BMM19}). 

Here, by referring to the continuous variable quantum perceptron model introduced in~\cite{BMM19}, we study the storage capacity of random classical binary patterns. The components of the patterns and their assigned output 
classification are taken to be independent and identically distributed (\textit{i.i.d.}) according to a probability with a bias $-1\leq \Min\leq 1$ for the patterns and $-1\leq \Mout\leq 1$ for their classification. Such a model 
admits the classical perceptron as a classical limit, thus enabling a direct comparison of the storage performances in the two cases. 
For classical perceptrons, simultaneously large biases for  patterns and output classification allow to {greatly enhance} the storage capacity, which diverges when $\Min=\Mout=m\to1$ \cite{G88,G87,GD88,AWS97}.
We {show} that this possibility is prevented at the quantum level.
Moreover, we {also} find that, when the biases $\Min$ and $\Mout$ are varied separately, the quantum storage capacity depends on both of them, {unlike} 
in the classical case, where the storage capacity is a function only of the output bias. 
However, also in the quantum setting, when the 
{output} correlations are maximal, that is when $\Mout\to 1$, the asymptotic behaviour is no more dependent on $\Min$, exactly as in the classical case. The dependence of the quantum storage capacity on $\Min$ 
in such a case  is through the velocity with which the limit behaviour is reached. Overall, the performances of the continuous quantum model remain below the classical ones.
These results thus corroborate those found in Ref.\cite{BGM22} with unbiased patterns. They confirm that, at the level of a 
simple, that is one-layer, quantum perceptron, the uncertainties brought about by pattern encoding via Gaussian states and homodyne measurements cannot be counteracted by linear super-positions of pattern states.


\section{A continuous quantum perceptron model}

The continuous variable model of a quantum perceptron proposed in Ref.\cite{BMM19}
is characterized by $N$ bosonic input modes and one bosonic output mode. The components $x_j$ of an input pattern $\xv\in\mathbb{R}^N$ are encoded by states of the form 
	\begin{equation}
		\vert \psi_j\rangle = \frac{1}{(2\pi \sigma^2_j)^{1/4}} \int_{-\infty}^{+\infty} \de q_j \  {  \exp\left(-\frac{(q_j - x_j)^2}{4 \sigma^2_j}\right)}\vert q_j\rangle,
		\label{eq:WavePacket}
	\end{equation}
	which are Gaussian weighted normalized super-positions of  pseudo-eigenstates $\vert q_j\rangle$ of position-like operators $q_j$, centered around the pattern components $x_j$ with widths $\sigma_j$. 
As a result, a pattern $\xv$ is encoded into
	\begin{equation}
		\vert \Psi\rangle=\bigotimes_{j=1}^N\vert \psi_j\rangle.
		\label{eq:Nwavepacket}
	\end{equation}
Such a state is then given as input to a quantum circuit which first operates with a series of independent squeezing
	operators
	\begin{equation}
		\label{Sq1}
		S_j(r_j)=\rme^{\rmi\,r_j\,(q_jp_j+p_jq_j)}\ ,\quad r_j\in\mathbb{R}\ ,\quad {\rm e}^{-2r_j}=w_j\ ,
	\end{equation} 
	where $p_j$ is a momentum-like operator conjugated to $q_j$ ($[q_j,p_j]=\rmi$) and $r_j$ is the squeezing parameter {implementing} 
 the weight $w_j$\footnote{In order to implement negative weights, a phase shift gate $\rme^{\rmi\frac{\pi}{2}(q_j^2+p_j^2)}$ is operated after the squeezing.}. Notice that 
 \begin{equation}
      S_j(r_j)\vert q_j\rangle=\sqrt{w_j}\,\vert w_jq_j\rangle\ .
 \end{equation}
	Then, the circuit {consists of} 
 entangling Controlled Addition gates $\CX$ on pairs of consecutive modes:
	\begin{equation}
		\label{eq:CX}
		{\CX_{j,j+1}}:= \exp\left(-\rmi\, q_{j} \otimes p_{j+1}\right)\ ,\qquad {\CX_{j,j+1}}\,\vert q_j, q_{j+1}\rangle=\vert q_j,q_j+q_{j+1}\rangle\ .
	\end{equation}
	{Their combined action on the attenuated multi-mode  position eigenstates 
 gives}
	\begin{equation}
		\begin{split}
			\vert w_1 q_1, w_2 q_2, \hdots,w_N q_N\rangle & \rightarrow \vert w_1 q_1, w_1 q_1+w_2 q_2, \hdots,w_N q_N\rangle  \rightarrow \hdots \\ 
			& \hdots \rightarrow \vert w_1 q_1,w_1 q_1+w_2 q_2, \hdots,\sum_{j=1}^{N} w_j q_j\rangle\ .
		\end{split}
	\end{equation}
In such a way, the amplitude associated to the last mode position eigenstate has the form of a Gaussian distribution centered around {$\wv\cdot\xv^\mu=\sum_{j=1}^Nw_jx^\mu_j$:
		\begin{equation}
			\psi_{\wv,\xv^{\mu}}(s)=\frac{1}{(2\pi\|\wv\|^2\,\sigma^2)^{1/4}}\exp\left(-\frac{(s-\wv\cdot \xv^\mu)^2}{4\|\wv\|^2\,\sigma^2}\right),\quad \|\wv\|^2=\sum_{j=1}^Nw^2_j\ ,
		\end{equation}
where, for sake of simplicity, we have set $\sigma_j^2=\sigma^2$ for all $j$ and thus encoded the input patterns by Gaussian states of the same width.}
Finally, homodyne detection operated on the last mode {position-like quadrature yields a value $s$ with 
probability density} 
		\begin{equation}\label{eq:Prob}
			P_{\wv,\xv^{\mu},\sigma}(s)=|\psi_{\wv,\xv^{\mu}}(s)|^2=\frac{1}{\sqrt{2\pi}\|\wv\|\sigma}\exp\left(-\frac{(s-\wv\cdot \xv^\mu)^2}{2\|\wv\|^2\sigma^2}\right).
		\end{equation}
\begin{rem}
With a slight modification of the {above protocol}, it is possible to obtain a description of such a continuous variable quantum perceptron as controlled unitary acting on the tensor product $\Hi=\Hi_{\mathrm{in}}\otimes \Hi_{\mathrm{out}}$ where $\Hi_{\mathrm{in}}$ is the Hilbert space of the $N$ bosonic modes encoding the input, while $\Hi_{\mathrm{out}}$ is the Hilbert space of the additional ancilla mode storing the output. Then, the present model can be connected with other models investigated in the literature, in particular~\cite{Beer}, where it was pointed out that a perceptron acting as a controlled unitary has as particular cases also the models considered in~\cite{Tor19,Cao}. Actually, the action of the continuous variable quantum 
perceptron here investigated can be described with the unitary
\begin{equation}
    U(\boldsymbol{w}):=\prod_{j=1}^N \CX_{j,\mathrm{out}}S_j(r_j),
\end{equation}
with $S_j(r)$ as in Eq.~\eqref{Sq1}, while $\CX_{j,\mathrm{out}}=\exp\left(-\rmi\, q_{j} \otimes p_{\mathrm{out}}\right)$ is the controlled addition gate involving the $j$-th bosonic mode of the input and the output mode. 
\end{rem}

Classically, the classification of a pattern $\xv^\mu$ as $\pm1$ by a weight vector $\wv$ is obtained by checking 
the sign of $\wv\cdot\xv^\mu$. Then, a correct classification relative to a prescribed target 
$\xi^\mu=\pm 1$ is obtained when $\xi^\mu\,\wv\cdot\xv^\mu\geq\kappa\|\wv\|$ where $\kappa$ is a stabilizing threshold. It renders the classification more robust against noise {affecting} the weights that, when $\kappa=0$,  might make $\xi^\mu\,\wv\cdot\xv^\mu$ {jump} from positive to negative values and vice versa.  In the case of the quantum perceptron model outlined above, a pattern $\xv^\mu$ is classified as $\xi^\mu=+1$ (resp. $\xi^\mu=-1$) if the measurement outcome is above the threshold $\kappa\|\wv\|$ (resp. below $-\kappa\|\wv\|$), while the pattern is not classified when the measurement outcome is in between $(-\kappa\|\wv\|,\kappa\|\wv\|)$.
Therefore, even when classically $\mathrm{sign}(\wv\cdot\xv^\mu-\kappa\|\wv\|)=+1$, quantumly, the pattern is classified as $-1$ if $s<-\kappa\,\|\wv\|$ and such  errors occur with probability density 
$P_{\wv,\xv
^{\mu},\sigma}(s)$.

Consequently, the inherent randomness due to the quantum encoding of the patterns is such that the correct classification
of pattern $\mu$ becomes a binary stochastic variable with probability distribution given by
	\begin{equation}\label{eq:rmu}
		{R_{\kappa,\sigma}(\wv,\xv^\mu,\xi^\mu)}={\int_{-\infty}^{+\infty}}  \de s\ P_{\wv,\xv^{\mu},\sigma}(s)\,
		\Theta\left(\xi^\mu\frac{s}{\|\wv\|}-\kappa\right),
	\end{equation}
where $\Theta(\cdot)$ denotes the Heaviside function. 	
Finally, an ancilla mode is appended to the initialized $N$ ones and its state changed according to the actual outcome 
of a suitable homodyne measurement. The result of the measurement can then be used, for instance, to implement the non-linear \textit{ReLu} activation function as in Ref.\cite{BMM19}.

One of the advantages of the continuous quantum model just presented is that it allows to recover the functioning of the classical perceptron when $\sigma\to 0$, i.e.
by encoding a pattern $\xv^\mu$ into the position-like pseudo-autokets $\vert x^\mu_1,x^\mu_2,\ldots,x^\mu_N\rangle$. 
Indeed, in this limit the Gaussian probability density in Eq.~\eqref{eq:rmu} becomes a Dirac delta centered around $\wv\cdot\xv^\mu$.


\section{Statistical Mechanics Derivation of the Storage Capacity}
\label{sec:storcap}

\subsection{Gardner's approach}
{According to Gardner’s statistical approach~\cite{G88}, the {optimal} storage {capacity of} a simple perceptron can be obtained from} the fraction of weights which correctly reproduces the desired input-output
relations normalized to the total volume of allowed vectors $\wv$. Indeed, the storage capacity is defined as the 
{\emph{critical} value $\alpha_c$ of the ratio 
\begin{equation}\label{alphadef}
\alpha\equiv\frac{p}{N},
\end{equation}
of the number of patterns $p$ to the dimension of the input space $N$ such that the storage condition  
\begin{equation}
    \label{storagecond}
    \xi^\mu\frac{\boldsymbol{w}\cdot\boldsymbol{x}^\mu}{\|\boldsymbol{w}\|}\geq \kappa
\end{equation}
{cannot be satisfied anymore.} 
}

In fact, by increasing the number of patterns, the volume of
vectors $\wv$ {realizing the condition \eqref{storagecond}} typically shrinks, and the relative volume of
{such} weights vanishes. Then, it is exactly the limit of vanishing relative volume
that leads to the storage capacity of the perceptron.

{We shall consider weights for which  $\|\wv\|^2=N$ so that their components are typically of order $1$.
Then,}  the fraction of weights $\boldsymbol{w}$ of length $\sqrt{N}$ in $\mathbb{R}^N$ that classify  $p$ binary patterns $\boldsymbol{x}^\mu\in\{+1,-1\}^N$, up to an error $\epsilon$, is given by:
	\begin{equation}
		\label{ref1}
V^Q_N(\xv^\mu,\xi^\mu,\kappa,\sigma,\epsilon) := \frac{1}{Z_{N}} \int_{\mathbb{R}^N}  \dd^N\boldsymbol{w} \ \delta(||\boldsymbol{w}||^2 - N) \prod_{\mu = 1}^{p} \Theta\Big({R_{\kappa,\sigma}(\wv,\xv^\mu,\xi^\mu)} - 1+\epsilon\Big).
	\end{equation}
where {the total volume of the space of the weights is}
	\begin{equation}
		\label{eq:NV}
{Z_{N}} = \int_{\mathbb{R}^N}^{} \dd^N\boldsymbol{w} \ \delta(||\boldsymbol{w}||^2 - N) 
		  \stackrel{N\gg1}{\simeq} \sqrt{\frac{(2\pi e)^N}{4\pi N}}\ , 
	\end{equation}
namely the volume of the sphere of radius $\sqrt{N}$ in $\mathbb{R}^N$. The relation to the classification of the pattern $\boldsymbol{x}^\mu$ is due to the fact that Eq.\eqref{eq:Prob} represents the probability distribution of the measurement outcomes of the quantum perceptron encoding the patterns 
$\boldsymbol{x}^\mu$ into Gaussian states of variance $\sigma$.
	
\begin{rem}
		\label{rem_error}
		The probability ${R_{\kappa,\sigma}(\wv,\xv^\mu,\xi^\mu)}$ depends on the {pattern} $\boldsymbol{x}^\mu$, on the {target classification} $\xi^\mu$, on the weights $\boldsymbol{w}$, on the threshold parameter $\kappa$ and on the Gaussian width $\sigma$.
		Therefore, the fraction of volume $V_N^Q$ depends on patterns, targets, threshold, width and also on the allowed statistical error $\epsilon$: when the width $\sigma$ vanishes, from the distributional limit
		\begin{equation}
\lim_{\sigma\to0}P_{\boldsymbol{w},\boldsymbol{{x}}^\mu,\sigma}(s)=\delta(s-\boldsymbol{w}\cdot\boldsymbol{x}^\mu)\ ,
		\end{equation}
		one recovers the expression of the fraction of weights  of the classical perceptron
		\begin{equation}
			\label{classV}
		{V^{C}_N(\boldsymbol
			{x}^\mu,\xi^\mu,\kappa)} = \frac{1}{Z_{N}} \int_{\mathbb{R}^N}^{}
			\dd^N\boldsymbol{w} \
			 \delta(\|\boldsymbol{w}\|^2 - N) \prod_{\mu = 1}^{p} 
			\Theta\left(\xi^\mu\frac{\boldsymbol{w}\cdot\boldsymbol{x}^\mu}{\|\boldsymbol{w}\|}-\kappa\right).
		\end{equation}
		Notice that this expression does not depend on the statistical error $\epsilon$ that needs to be introduced in the quantum setting. Indeed, in this latter case, the measured parameter $s$ is statistically distributed around the classical scalar product $\boldsymbol{w}\cdot\boldsymbol{x}^\mu/\|\boldsymbol{w}\|$. Therefore, ${R_{\kappa,\sigma}(\wv,\xv^\mu,\xi^\mu)}$ cannot be equal to $1$ unless the Gaussian distribution becomes a Dirac delta peaked around it. Note that the statistical error $\epsilon$ is an upper bound to the perceptron allowed errors. {The value $\epsilon=1/2$ for the bound to the errors is a particular one: in such a case weights can provide classifications with equal probability of being right or wrong. Therefore, at $\epsilon=1/2$, weights are selected without further constraints beside the classical ones. Then, as far as the storage capacity is concerned,  the quantum perceptron is expected to behave classically 
  at $\epsilon=1/2$, in spite of the quantum pattern encoding.} 
	\end{rem}

	In analogy with the partition function of statistical mechanics, we take ${\log 
	V^Q_N(\xv^\mu,\xi^\mu,\kappa,\sigma,\epsilon)}$ as the relevant quantity, since it has the important property of being \textit{self-averaging}, i.e. its average ${\braket{\log V^Q_N(\xv^\mu,\xi^\mu,\kappa,\sigma,\epsilon)}}$ is a good representative of its typical behaviour for random choices of input patterns and targets~\cite{Shcherbina2003,Tal,Engel}. In particular, this average will be computed considering the components of the input patterns as well as targets, to be binary stochastic variables distributed according to 
	\begin{equation}
		\label{probpatt}
		\Pr(x^{\mu}_j=\pm1)=\frac{1\pm \Min}{2}\ ,\qquad \Pr(\xi^\mu=\pm 1)=\frac{1\pm \Mout}{2}\ .
	\end{equation}
     The parameters $-1\leq \Min, \Mout \leq 1$ measure the bias between the binary values of patterns and targets, 
     respectively, and thus of their correlations. The smaller the bias is, the greater the independence of their two 
     possible values.
	
Following the classical approach by Gardner we will derive a critical value $\alpha_c^Q$ 
such that for $\alpha<\alpha_c^Q$ we obtain a finite value (potentially vanishing) for the limit
\begin{equation}
\label{eq:aveLogV1}
	\lim_{\stackrel{N,p\rightarrow \infty}{{  p/N=\alpha}}}\frac{\langle \log V^Q_N(\xv^\mu,\xi^\mu,\kappa,\sigma,\epsilon)) \rangle}{N}\ ,
\end{equation}
while for $\alpha> \alpha_c^Q$:
\begin{equation}
\label{eq:aveLogV2}
	\lim_{\stackrel{N,p\rightarrow \infty}{{  p/N=\alpha}}}\frac{\langle \log V^Q_N(\xv^\mu,\xi^\mu,\kappa,\sigma,\epsilon) \rangle}{N} = - \infty \ .
\end{equation}
\subsection{Replica method and saddle point equations}
The quenched average appearing in the limit \eqref{eq:aveLogV1} can be computed by means of the replica-trick \cite{G88,MezardParisi,Castellani}:
	\begin{equation}
		\label{eqf7}
		\braket{\log{V^Q_N(\xv^\mu,\xi^\mu,\kappa,\sigma,\epsilon)}} = \lim_{n \rightarrow 0} 
\frac{\braket{{[V^Q_N(\xv^\mu,\xi^\mu,\kappa,\sigma,\epsilon)]}^n} - 1}{n}
  =\lim_{n\rightarrow 0} \frac{\log\braket{{[V^Q_N(\xv^\mu,\xi^\mu,\kappa,\sigma,\epsilon)]}^n}}{n}.
	\end{equation}
 The relevant quantity $\braket{{[V^Q_N(\xv^\mu,\xi^\mu,\kappa,\sigma,\epsilon)]}^n}$ involves $n$ replicas indexed by the subscript $\gamma$:
	\begin{equation}
		\label{VmeanText}
		\langle{[V^Q_N(\xv^\mu,\xi^\mu,\kappa,\sigma,\epsilon)]}^n\rangle	= \frac{1}{Z_N^n}{\int_{\mathbb{R}^{nN}}\dd^{nN}\boldsymbol{W}~\delta_N^{(n)}(\|\boldsymbol{W}\|^2)}\,\left\langle\prod_{\gamma=1}^n\prod_{\mu = 1}^{p} \Theta(R^\mu_{\gamma} - 1 + \epsilon)\right\rangle_{\boldsymbol{x},\xi},
	\end{equation}	
 where, {for the sake of compactness, we introduced the symbols $\boldsymbol{W}=(\boldsymbol{w}_1,\ldots,\boldsymbol{w}_n)\in\mathbb{R}^{nN}$ and
 \begin{equation}
 \label{aux0}
\dd^{nN}\boldsymbol{W}\equiv\prod_{\gamma=1}^n\dd^N\boldsymbol{w}_\gamma\ ,\quad 
 \delta_N^{(n)}(\|\boldsymbol{w}\|^2)\equiv\prod_{\gamma=1}^n\delta(\|\boldsymbol{w}_\gamma\|^2-N)\ .
 \end{equation} 
 Moreover,} {in Eq.\eqref{VmeanText},} it is made explicit that the mean value is computed with respect to the patterns 
 $\boldsymbol{x}^\mu$ and targets $\xi^\mu$, $\mu=1,\ldots,p$. 
		
The lengthy calculations of the mean value in Eq.\eqref{VmeanText} by means of the replica-symmetric ansatz and of the saddle point approximation are reported in Appendix \ref{appA}. 

\begin{figure}
\centering 
\includegraphics[width=0.49\textwidth]{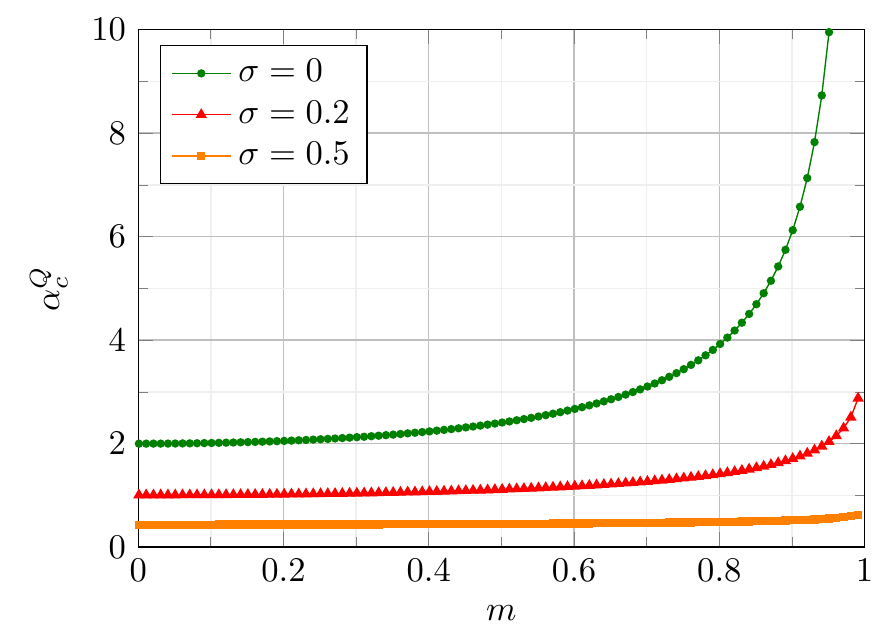}
\includegraphics[width=0.49\textwidth]{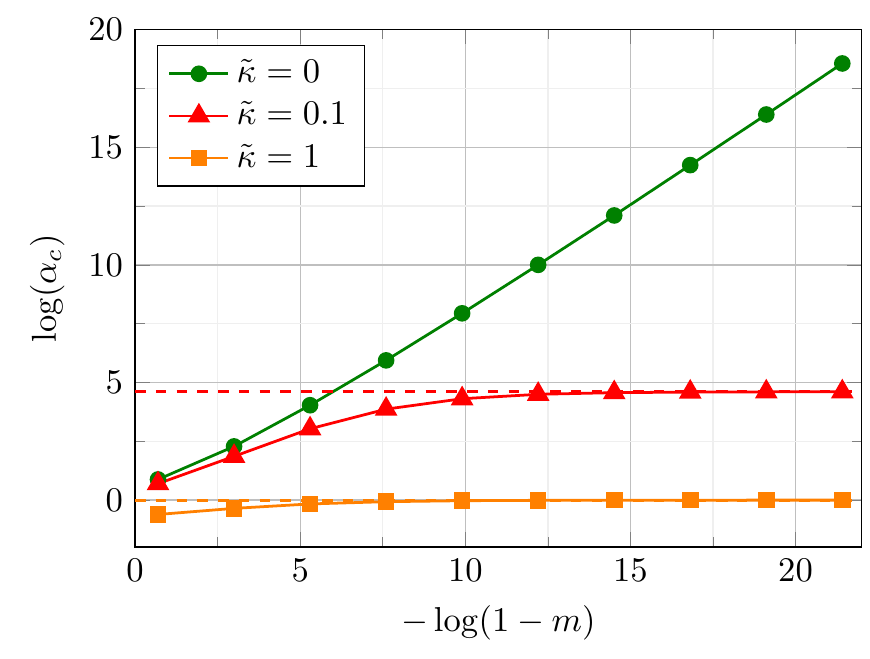}
\caption{(Left) $\alpha_c^{Q}$ plotted against $m=\Min=\Mout$. Increasing the value of the bias $|m|$ always increases the storage capacity. (Right) Limit $\alpha_c^{Q}$ for $m_{\mathrm{out}}=m_{\mathrm{in}}=m\rightarrow 1$, for several values of $\tilde{\kappa}$. The case $\tilde{\kappa}=0$ corresponds to the classical case with $\kappa=0$, where the storage capacity diverges in the limit. In the quantum scenario ($\tilde{\kappa}>0$) the divergence is suppressed. The dashed lines correspond the asymptotic value $\alpha_c\sim 1/\tilde{\kappa}^2$}
\label{fig:alpha_m}
\end{figure}

{The replica method introduces several order parameters, the most important one being the average overlap of two randomly chosen weights $\wv_\gamma$ and $\wv_\delta$ in different replicas:
\begin{equation}
\label{orderparameter}
q_{\gamma \delta}=\frac{1}{N}\sum_{j=1}^{N} w_j^\gamma w_j^\delta\ .
    \end{equation}
In the replica symmetric ansatz it is assumed that for the solution of the saddle point equations, the average overlap is the same for each pair of replicas, i.e. $q_{\gamma\delta}=q$ for all $\gamma\neq \delta$. Notice that by increasing the ratio $p/N$, the number of weights satisfying~\eqref{storagecond} diminishes, hence their average 
overlap increases. The critical value of 
$\alpha_c$ both in the classical and the quantum scenario is then obtained in the limit of maximal overlap $q\to1$.


Eventually, one arrives at the following equation that must be satisfied by
the critical ratio $\alpha_c^Q$ of number of patterns to weight dimension, which according to \eqref{alphadef} defines the quantum storage capacity:
\begin{equation}
\label{eq:alpha1}
\alpha_c^Q\Bigg[ \frac{1+\Mout}{2} \int_{a_-(M)}^{+\infty}
		 \DD x\,(x-a_-(M))^2\,+\,\frac{1-\Mout}{2}\int_{a_+(M)}^{+\infty}
		\DD x\,(x-a_+(M))^2 \Bigg] = 1.
\end{equation}
In the above expression,
\begin{equation}
\label{apm}
a_\pm(M)\equiv-\frac{\tilde{\kappa}\pm \Min M}{\sqrt{1-\Min^2}}\ ,\quad \DD x \equiv \frac{\dd x}{\sqrt{2\pi}} e^{-x^2/2}\ ,
\end{equation}
where
\begin{equation}\label{eq:kappat}
    \tilde{\kappa}:=\kappa+\sigma\Phi^{-1}(1-\epsilon),
\end{equation}
and $\Phi^{-1}$ is the inverse function of 
\begin{equation}
\label{erfunct}
\Phi(x)\equiv\int_{-\infty}^x \DD u =\frac{1+{\rm erf}(x/\sqrt{2})}{2} \quad \hbox{with}\quad\hbox{ erf}(x):=\frac{2}{\sqrt{\pi}}\int_0^x \dd u\,{\rm e}^{-u^2}\ ,
\end{equation}	
while the quantity $M$ satisfies
\begin{equation}
\label{eq:alpha2}
		(1+\Mout) \int_{a_-(M)}^{+\infty} \DD x\,(x-a_-(M))=(1-\Mout) \int_{a_+(M)}^{+\infty}\DD x \, (x-a_+(M))\ .
\end{equation}
Thus, in order to compute $\alpha_c^Q$ from \eqref{eq:alpha1}, one 
has to first solve \eqref{eq:alpha2} in terms of $M$.

\begin{rem}
   Notice that when $\Min=0$, that is when the patterns are unbiased, we have $a_+(M)=a_-(M)$ so that Eq.~\eqref{eq:alpha2} can be satisfied only for $\Mout=0$, and the storage capacity is fixed by~\eqref{eq:alpha1} only, which coincides with the expression found in~\cite{BGM22}. This is due to the fact that a perceptron cannot {match 
   unbiased patterns with biased classifications}. In fact, considering the mapping realized by a perceptron with weights $\boldsymbol{w}$, one finds that for a random input $\boldsymbol{x}\in\R^N$, with independent components distributed according to $\Pr(x_j= \pm 1)=1/2$ for each $j=1,\dots, N$, the distribution of the output $\sigma$ is given by:
\begin{equation}
    \Pr(\sigma=+ 1)=\Pr(\boldsymbol{w}\cdot\boldsymbol{x}\geqslant 0)=\Pr(\boldsymbol{w}\cdot\boldsymbol{x}\leqslant 0)=\Pr(\sigma=- 1),
\end{equation}
which implies $\Pr(\sigma=\pm 1)=1/2$ (note that $\Pr(\boldsymbol{w}\cdot \boldsymbol{x}=0)$ for all $\boldsymbol{w}\in \R^N$ except for those belonging to a set with zero Lebesgue measure on the sphere with radius $\sqrt{N}$).
\end{rem}

In practice, the combined effects of quantum pattern encodings and measurements is to replace the classical 
stabilizing threshold $\kappa$ in $\tilde{\kappa}$ defined in \eqref{eq:kappat}. Then, the classical storage capacity obtains not only by eliminating the errors due to quantum pattern encoding, that is by letting $\sigma\rightarrow 0$, but also, {confirming the argument in Remark~\ref{rem_error},
when $\sigma\neq 0$, so that the pattern encoding is not sharp and carries quantum fuzziness; however, $\epsilon=1/2$ so that $\Phi^{-1}(1-\epsilon)=0$ and $\tilde{\kappa}=\kappa$.}

\section{Results}
\begin{figure}[]
    \centering
    \includegraphics[width=0.325\textwidth]{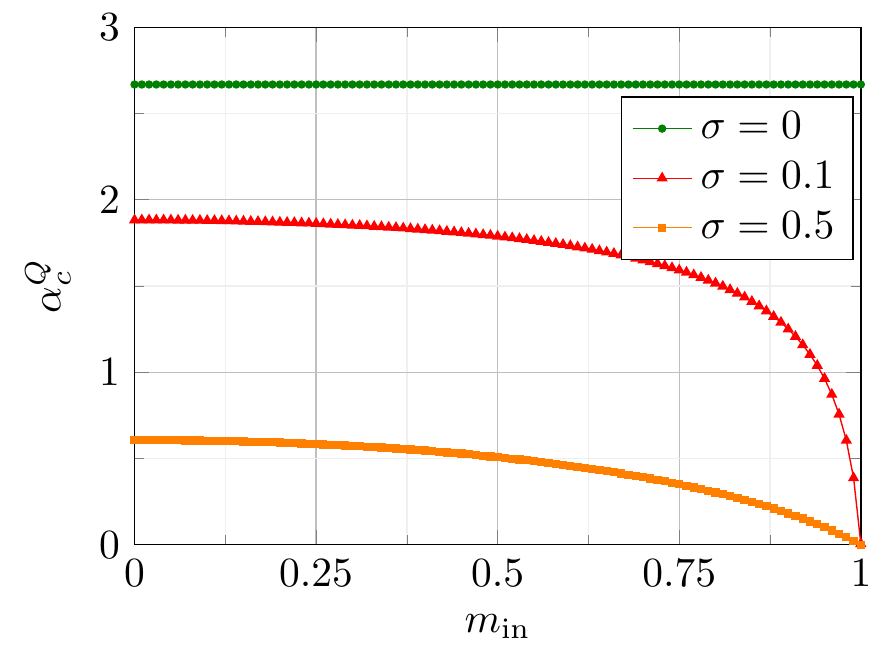}~
    \includegraphics[width=0.325\textwidth]
{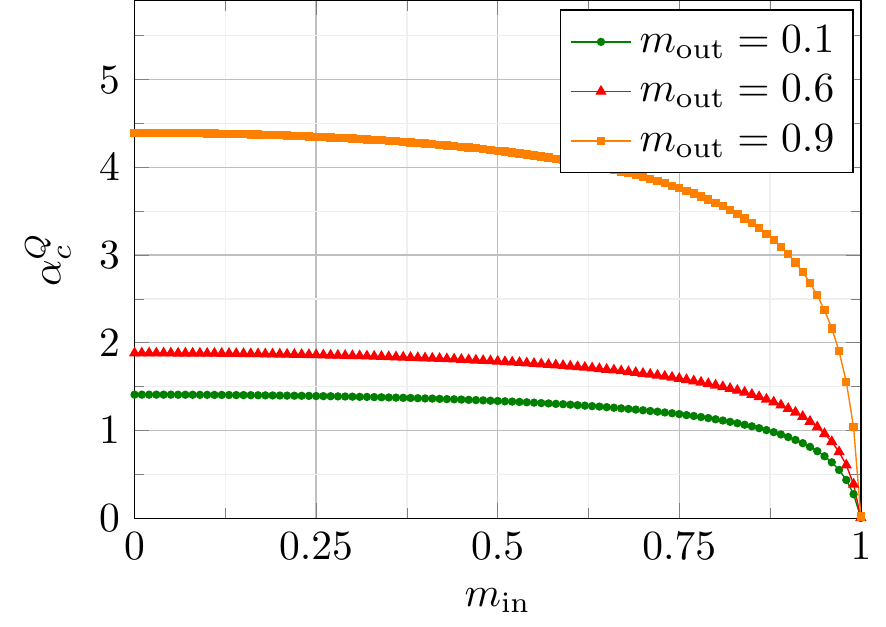}~
    \includegraphics[width=0.325\textwidth]{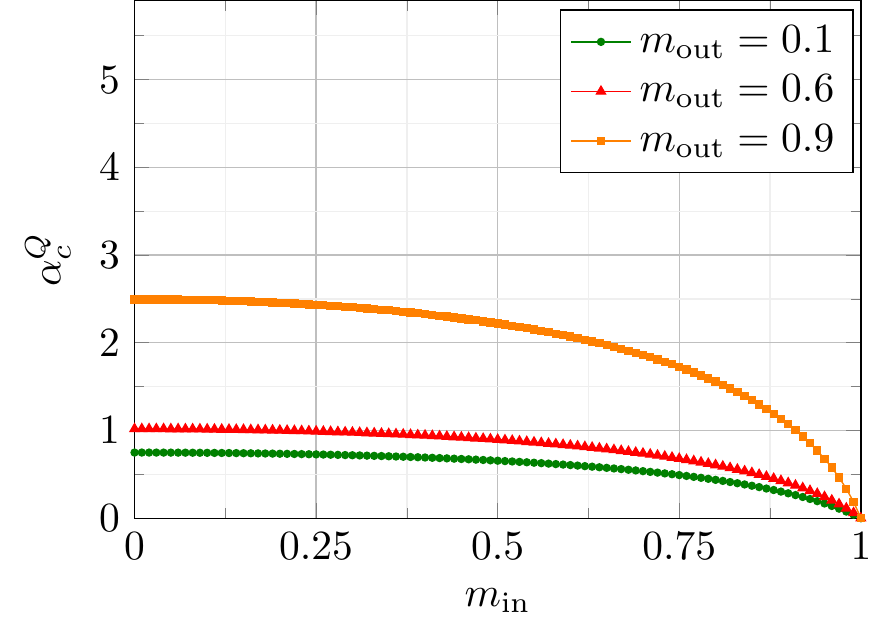}
  \caption{Storage capacity $\alpha_c^{Q}$ vs $\Min$, for $\kappa=0$, and $\epsilon=0.01$. Only values $0<\Min<1$ are shown here, but the results are symmetric with respect to $\Min=0$. (Left) The bias on the target classification is fixed to $\Mout=0.6$, the shown curves corresponding to different values of $\sigma$. In the classical case ($\sigma=0$) $\alpha_c^{Q}$ does not depend on $\Min$. In the quantum case $(\sigma>0$), increasing the value of $|\Min|$ decreases $\alpha_c^{Q}$. Furthermore, increasing $\sigma$ always decreases the storage capacity. (Center) The value $\sigma=0.1$ is fixed, the shown curves corresponding to different values of $\Mout$. Increasing the value of $|\Mout|$ always increases the storage capacity (Right) As before, with $\sigma=0.3$, {showing the lowering of the perceptron performance with increasing quantum fuzziness in the pattern encoding.}
  \label{fig:alpha_MoutMin}}
\end{figure}

\begin{figure}[] 
    \centering  \includegraphics[width=0.49\textwidth]{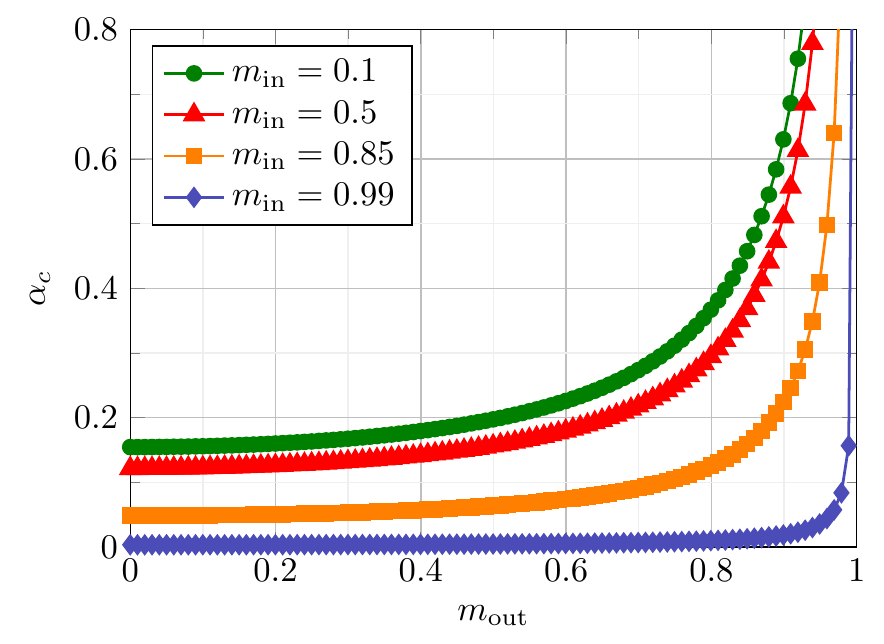}
    \includegraphics[width=0.49\textwidth]{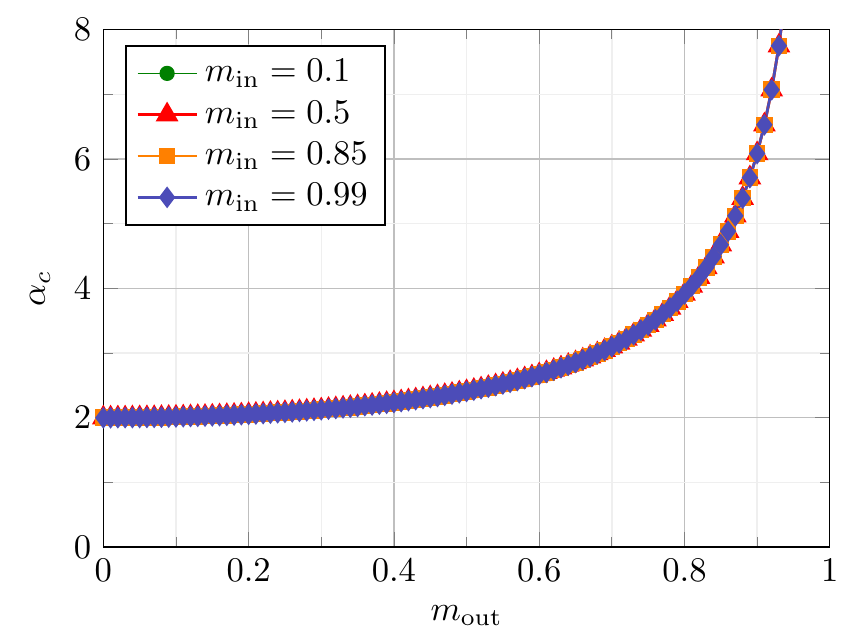}
   \caption{(Left) Storage capacity $\alpha_c^{Q}$ vs $\Mout$ for different values of $\Min$, $\kappa=0$, $\sigma=1$ and $\epsilon=0.01$. Increasing $|\Mout|$ always increases $\alpha_c^{Q}$, while increments of $|\Min|$ have the opposite effect. There is a divergence for $|\Mout|\rightarrow 1$ for each value of $\Min$, although higher values of $|\Mout|$ are required to observe the divergence if $|\Min|$ is increased. (Right) Storage capacity in the classical limit, obtained for $\sigma=0$ (all the values of the other parameters are unchanged). All the curves corresponding to different values of $\Min$ collapse into each other since in this case there is no dependence on $\Min$ (recall that here $\kappa=0$.}
  \label{fig:alpha_MinMout} 
\end{figure}

The numerical results obtained by solving equations \eqref{eq:alpha1} and \eqref{eq:alpha2} for several values of $\Min$, $\Mout$ and $\tilde{\kappa}$ are shown in Figure~\ref{fig:alpha_MoutMin} and Figure~\ref{fig:alpha_MinMout}. Since the storage capacity depends solely on $\widetilde{\kappa}=\kappa+\sigma \Phi^{-1}(1-\epsilon)$, we kept fixed the values $\epsilon=0.01, \kappa=0$ and considered different values of $\sigma$, which also allows us to recover the classical limit for $\sigma=0$. A striking feature which distinguishes the quantum perceptron from the classical one is the dependence of the storage capacity on the bias $\Min$, which is not present in the classical case with zero stability $\kappa=0$ (see the left panel of  Figure~\ref{fig:alpha_MoutMin}). More precisely, as soon as $\sigma>0$, increasing the value of $|\Min|$ while keeping fixed the value of $\Mout$ always decreases the storage capacity $\alpha_c^{Q}$. On the other hand, a common feature with the classical case is the divergence of the storage capacity when $\Mout\rightarrow 1$, for each fixed value of $\Min$ (see Figure~\ref{fig:alpha_MinMout}) The asymptotic behaviour in this limit (see Appendix~\ref{appB} for the analytic derivation) is given by (see also Figure~\ref{fig:limitAlpha}):
\begin{equation}
\label{asymp0}
    \alpha_{c}^Q \simeq -\frac{1}{(1-\Mout)\log(1-\Mout)}\ ,
\end{equation}
confirming the result obtained for the classical scenario.
\begin{figure}
    \centering
    \includegraphics[width=.6\textwidth]{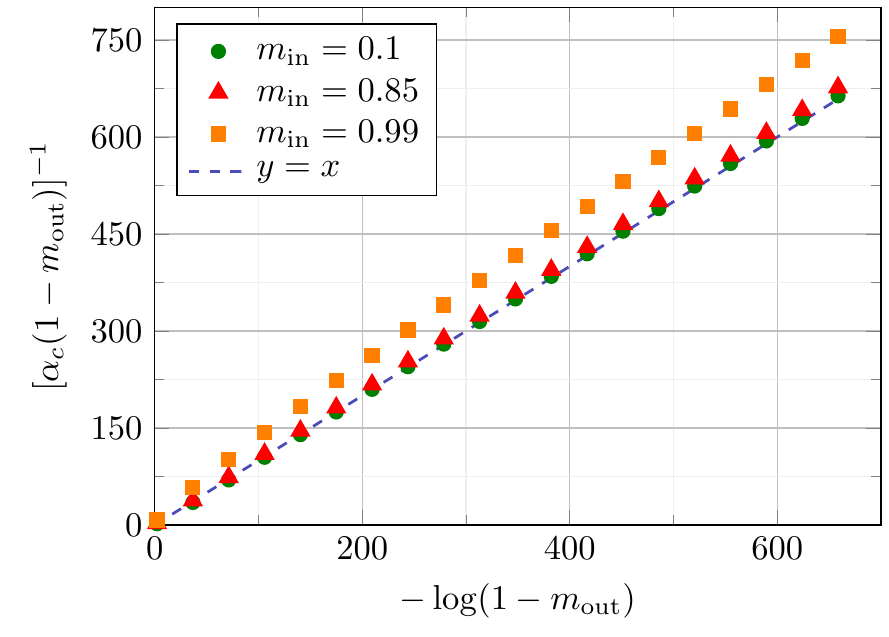}
    \caption{Limit $\alpha_c$ for $m_{\mathrm{out}}\rightarrow 1$, showing that the convergence to the asymptotic behaviour \eqref{asymp0}, represented here by the line $y=x+q$, holds for several values of $m_{\mathrm{in}}$, although the onset of the asymptotic behaviour appears later for $m_{\mathrm{in}}$ close to one. }
    \label{fig:limitAlpha}
\end{figure}
\begin{rem}
An interesting feature emerging by differentiating the bias of the patterns $\Min$ from the bias of target classification $\Mout$ is that even if the {quantum} storage capacity depends on $\Min$ for each fixed $0\leqslant \Mout\leqslant 1$, the asymptotic behaviour 
when $\Mout\to1^-$ does not depend on the pattern biases $\Min$. 
\end{rem}

Even if the asymptotic behaviour \eqref{asymp0} does not show a dependence on the patterns bias $\Min$, one can see from Figure~\ref{fig:alpha_MinMout} and Figure~\ref{fig:limitAlpha} that as the input bias $\Min$ increases, higher values of $\Mout$ are required to observe the asymptotic behaviour for the quantum perceptron. This is in contrast with the behaviour of the classical perceptron, where there is no dependence at all on $\Min$ (see again Figure~\ref{fig:alpha_MinMout}}). In other words, values of $\Min$ closer to $1$ slow down the attainment of the asymptotic behaviour in the quantum case, which motivates the investigation of the joint limit $\Min=\Mout=m\rightarrow 1$. The results obtained (see Figure~\ref{fig:alpha_m}) show another striking difference between the classical and the quantum perceptron, that is, while classically the storage capacity diverges when $m\rightarrow1$, in the quantum case this divergence is suppressed. In particular, from the analytic asymptotic expressions (see Appendix~\ref{appC}) we obtain that the asymptotic behaviour in the quantum scenario reads
\begin{equation}
    \alpha_c^{Q}\sim \frac{1}{\tilde{\kappa}^2},
\end{equation}
which is finite for all values of $\sigma>0$, $0\leq \epsilon<1/2$, while we recover the classical divergence (for $\kappa=0$) in the classical limit $\sigma=0$.

\begin{rem}\label{rmk:sep}
Figure \ref{fig:alpha_MoutMin} shows that for $\kappa=0$ the separation between curves corresponding to different values of 
$\Mout$ is reduced in the quantum regime. This is in contrast to what happens for curves corresponding to different values 
of $\Min$, as shown in Fig.\ref{fig:alpha_MinMout}.
\end{rem}

\section{Discussion and conclusion}

Summarizing, we studied the storage capacity of the continuous variable model of quantum perceptron presented in Ref. \cite{BMM19} in the presence of a bias in the distribution of the patterns and their corresponding classifications. Besides the advantage of allowing an almost entirely analytical study, such a model admits the classical perceptron as a classical limit, thus allowing for a direct comparison of the
storage performances in the two cases. 
We found that the additional randomness introduced in the quantum model gives rise to an effective increment in the stability parameter used in Gardner's statistical approach, $\kappa\rightarrow \tilde{\kappa}$, which gives rise to  several peculiar features that are not observed in the classical case with zero stability. 

For instance, the possibility of indefinitely enhancing the storage capacity by increasing the bias of the patterns and their classifications is  prevented at the quantum level.
Moreover, in contrast to the classical case, when the bias of the patterns and the bias in their corresponding classifications are varied separately, a dependence of the storage capacity on the input patterns’ bias appears even when the stability parameter reduces to zero. 
Overall, however, the performance of the quantum perceptron model remains below that of the classical one. 
This is likely due to the fact that the considered quantum model introduces two sources of randomness: one due to the encoding 
of patterns by means of non-zero width Gaussian states and another one due to the final measurement operation implementing 
the classical non-linear activation function. 
It is worth stressing that the modification in the effective threshold $\tilde{\kappa}-\kappa=\sigma\Phi^{-1}(1-\epsilon)$ contains both the contribution of randomness coming from the width $\sigma$ of the Gaussian encoding of the patterns, and the statistical error due to the quantum measurement $\epsilon$: as a consequence, the worsening of the quantum storage capacity with respect to the classical one cannot be ascribed to only one of them.
{These results clearly point to the necessity of considering multi-layer quantum perceptrons to hope for quantum advantages of {the} sort coming  from linear superpositions and entanglement.}

\section*{Acknowledgements}
{F.B.,} G.G. and S.M. acknowledge financial support from PNRR MUR project PE0000023-NQSTI. {S.M. also acknowledges financial support from the European Union’s
Horizon 2020 research and innovation programme, 
under grant agreement QUARTET No 86264.}

\appendix

\section{Derivation of the main result}
\label{appA}	
{In order to extract the large $N$ behavior of}  
$\braket{{[V^Q_N(\xv^\mu,\xi^\mu,\kappa,\sigma,\epsilon)]}^n}$ in Eq.\eqref{Vmean}, we recast it as
	\begin{equation}
		\label{Vmean}
		\langle {[V^Q_N(\xv^\mu,\xi^\mu,\kappa,\sigma,\epsilon)]}^n\rangle	=\frac{1}{Z^n_N}\int_{\mathbb{R}^{nN}} \dd^{nN}\boldsymbol{W}~\delta_N^{(n)}(\|\boldsymbol{W}\|^2)\,{A(\boldsymbol{W})}\ ,
  \end{equation}
with
{\begin{equation}
 \label{Vmeanaux1}
 A(\boldsymbol{W}):=\left\langle\prod_{\mu = 1}^{p}\prod_{\gamma = 1}^{n} \Theta({R_{\kappa,\sigma}(\wv_\gamma,\xv^\mu,\xi^\mu)} - 1 + \epsilon)\right\rangle_{\boldsymbol{x},\xi} \ .
 \end{equation}
 }
Using the distributional expression
	\begin{equation}
		\label{eq:distTheta}
		\Theta(x - u)=\int_u^{+\infty} \dd\lambda \ \delta(\lambda - x)   = \int_u^{+\infty} \dd\lambda \
		{\int_{-\infty}^{+\infty}}  \frac{\dd y}{2\pi}  \ \ee^{\ii y(\lambda - x)},
	\end{equation}
	we express the Heaviside function in~\eqref{Vmeanaux1} as
	\begin{equation}
		\label{Dirac}	
		\Theta({R_{\kappa,\sigma}(\wv_\gamma,\xv^\mu,\xi^\mu)} - 1 + \epsilon) = 
		\int_{1-\epsilon}^{+\infty} \dd \lambda_{\gamma}^{\mu} \ {\int_{-\infty}^{+\infty}}
		\frac{\dd y_{\gamma}^{\mu}}{2\pi} \ 
		\ee^{\ii y_{\gamma}^{\mu}(\lambda_{\gamma}^{\mu} -{R_{\kappa,\sigma}(\wv_\gamma,\xv^\mu,\xi^\mu)} )}, 
	\end{equation}	
	so that 
	{
 \begin{equation}\label{Amean}
		A(\boldsymbol{W})=\frac{1}{(2\pi)^{nP}}\int_{[1-\epsilon,+\infty)^{nP}}\hskip -1cm\dd^{nP}\boldsymbol{\Lambda}
		{\int_{\R^{nP}}} \dd^{nP}\boldsymbol{Y}\ 
		\ee^{\ii \sum_{\mu=1}^p\sum_{\gamma=1}^ny_{\gamma}^{\mu}\lambda_{\gamma}^\mu}\, B(\boldsymbol{Y},\boldsymbol{W})\ ,
  \end{equation}
  where we introduced the symbols
  \begin{equation}
      \label{aux01}
      \dd^{nP}\boldsymbol{\Lambda}\equiv\prod_{\mu = 1}^{p}\prod_{\gamma = 1}^{n}
			\dd \lambda_{\gamma}^{\mu}\ ,\ \dd^{nP}\boldsymbol{Y}\equiv\prod_{\mu = 1}^{p}\prod_{\gamma = 1}^{n}
			\dd y_{\gamma}^{\mu}\ ,
  \end{equation}
  and the quantity
  \begin{equation}
  \label{Ameanaux1}
B(\boldsymbol{Y},\boldsymbol{W})\equiv\left\langle\prod_{\gamma=1}^n\prod_{\mu=1}^p~\ee^{-\ii y_{\gamma}^{\mu} {R_{\kappa,\sigma}(\wv_\gamma,\xv^\mu,\xi^\mu)} }\right\rangle_{\boldsymbol{x},\xi}\ .
	\end{equation}}
To proceed, it is convenient to use (see~\eqref{erfunct})
$$
\Phi(x)\equiv\frac{1}{\sqrt{2\pi}}\int_{-\infty}^x \dd u \ e^{-u^2/2}
$$
and rewrite~\eqref{eq:rmu} as 
	\begin{equation}
		{R_{\kappa,\sigma}(\wv_\gamma,\xv^\mu,\xi^\mu)} = \int_{-\infty}^{+\infty}
		\dd s \ 
		P_{\boldsymbol{w}_{\gamma},\boldsymbol{{x}}^\mu,\sigma}(s) \ \Theta\left(\xi^{\mu} \frac{s}{||\boldsymbol{w}||} -\kappa\right)  =\Phi
		\left(-\frac{\kappa}{\sigma} +\xi^{\mu}\frac{\boldsymbol{x}^{\mu}\cdot\boldsymbol{w}_\gamma}{||\boldsymbol{w}_{\gamma}||\sigma}\right).
	\end{equation}
Then, using the exponential representation of the Dirac delta and the statistical independence of patterns and targets with different indices, one gets
	\begin{align}
		\nonumber
		B(\boldsymbol{Y},\boldsymbol{W})&=\left\langle\prod_{\mu = 1}^{p}\prod_{\gamma = 1}^{n}\ee^{-\ii y_{\gamma}^{\mu} \Phi
		\left(-\frac{\kappa}{\sigma} +\xi^{\mu}\frac{\boldsymbol{x}^{\mu}\cdot\boldsymbol{w}_\gamma}{||\boldsymbol{w}_{\gamma}||\sigma}\right) }\right\rangle_{\boldsymbol{x}^\mu,\xi^{\mu}}\\
		\nonumber
		&=\left\langle\prod_{\mu = 1}^{p}\prod_{\gamma = 1}^{n}\int_{\R}
		\dd\eta^\mu_\gamma \
		 \delta\left(\eta^\mu_\gamma+\frac{\kappa}{\sigma} -\xi^{\mu}\frac{\boldsymbol{x}^{\mu}\cdot\boldsymbol{w}_\gamma}{||\boldsymbol{w}_{\gamma}||\sigma}\right)\, \ee^{-\ii y_{\gamma}^{\mu} \Phi\left(\eta^\mu_\gamma\right) }\right\rangle_{\boldsymbol{x},\xi}\\
		\nonumber
		&={\frac{1}{(2\pi)^{nP}}\int_{\mathbb{R}^{2nP}}\dd^{nP}\boldsymbol{\Omega}\dd^{nP}\boldsymbol{H}~
  C(\boldsymbol{\Omega},\boldsymbol{W})}\\
\label{Bmean}
		&{\qquad \qquad \qquad  \qquad 
		\times\exp\Bigg[-i\sum_{\mu=1}^p\sum_{\gamma = 1}^{n}\left(\eta^\mu_\gamma+\frac{\kappa}{\sigma}\right)\omega_{\gamma}^{\mu}-i\sum_{\mu=1}^p\sum_{\gamma = 1}^{n}y^\mu_\gamma \Phi(\eta^\mu_\gamma)\Bigg]}\ ,
	\end{align}
 with
 \begin{equation}   \label{Bmean1} C(\boldsymbol{\Omega},\boldsymbol{W}):=\prod_{\mu=1}^p\left\langle\exp\left(\ii\xi^{\mu}\sum_{\gamma = 1}^{n}\omega^{\mu}_{\gamma} \frac{\boldsymbol{x}^\mu\cdot\boldsymbol{w}_\gamma}{\sigma\sqrt{N}}\right) \right\rangle_{\boldsymbol{x},\xi}\ ,
 \end{equation}{where $\boldsymbol{\Omega}=\{\omega^\mu_\gamma\}_{\mu,\gamma}$, $\boldsymbol{H}=\{\eta^\mu_\gamma\}_{\mu,\gamma}$
and
\begin{equation}
\label{OH}
\dd^{nP}\boldsymbol{\Omega}\dd^{nP}\boldsymbol{H}=\prod_{\mu= 1}^{P}\prod_{\gamma= 1}^{n}\dd\omega_{\gamma}^{\mu}\dd\eta^{\mu}_{\gamma}\ .
\end{equation}
}
Since the patterns have components $x^\mu_j=\pm 1$ which are statistically independent and identically distributed, using~\eqref{probpatt}, one computes  the mean over the patterns in {$C(\boldsymbol{\Omega},\boldsymbol{W})$} as follows:
	\begin{align}
		\nonumber{C(\boldsymbol{\Omega},\boldsymbol{W})} &= \prod_{\mu=1}^p\prod_{j = 1}^N \Big\langle\exp\Bigg(\ii\xi^{\mu}\sum_{\gamma = 1}^{n}\omega^{\mu}_{\gamma} \frac{w_{\gamma,j} x^{\mu}_j}{\sigma\sqrt{N}}\Bigg) \Big \rangle_{x_j^{\mu},\xi^\mu}\\
		\nonumber &= \prod_{\mu=1}^p\prod_{j = 1}^N \frac{1}{2} \Bigg\langle \exp\Bigg(\ii\xi^{\mu}\sum_{\gamma = 1}^{n} \frac{w_{\gamma,j} \omega^{\mu}_{\gamma}}{\sigma\sqrt{N}}\Bigg) + \exp\Bigg(-\ii \xi^{\mu}\sum_{\gamma = 1}^{n} \frac{ w_{\gamma,j} \omega^{\mu}_{\gamma}}{\sigma\sqrt{N}}\Bigg)\\
		\nonumber
		&\qquad+\Min\Bigg( \exp\Bigg(\ii\xi^{\mu}\sum_{\gamma = 1}^{n} \frac{w_{\gamma,j} \omega^{\mu}_{\gamma}}{\sigma\sqrt{N}}\Bigg) - 
		\exp\Bigg(-\ii \xi^{\mu}\sum_{\gamma = 1}^{n} \frac{w_{\gamma,j} \omega^{\mu}_{\gamma}}{\sigma\sqrt{N}}\Bigg)\Bigg)
		\Bigg\rangle_{\xi^\mu}
		\\
		\nonumber &=\prod_{\mu=1}^p\prod_{j = 1}^N\left\langle \cos{\Bigg(\xi^{\mu}\sum_{\gamma = 1}^{n} \frac{w_{\gamma,j} \omega^{\mu}_{\gamma}}{\sigma\sqrt{N}}\Bigg)}+\ii \Min\sin{\Bigg(\xi^{\mu}\sum_{\gamma = 1}^{n} \frac{ w_{\gamma,j} \omega^{\mu}_{\gamma}}{\sigma\sqrt{N}}\Bigg)}\right\rangle_{\xi^\mu}\\
		\nonumber &=\prod_{\mu=1}^p\prod_{j = 1}^N \left\langle\exp{\Bigg(\log \left[\cos{\Bigg(\xi^{\mu}\sum_{\gamma = 1}^{n} \frac{ w_{\gamma,j} \omega^{\mu}_{\gamma}}{\sigma\sqrt{N}}\Bigg)}+\ii \Min\sin{\Bigg(\xi^{\mu}\sum_{\gamma = 1}^{n} \frac{ w_{\gamma,j} \omega^{\mu}_{\gamma}}{\sigma\sqrt{N}}\Bigg)}\right]\Bigg)}\right\rangle_{\xi^\mu}.
	\end{align}		
When $N\gg 1$, the leading order expansion in $1/N$ of each factor in the product over the $\mu$ index reads			
	\begin{align}		
			\nonumber&\prod_{j = 1}^N 
\left\langle\exp\Bigg( \log \Bigg[1+\ii \Min\Bigg(\xi^{\mu}
   \sum_{\gamma = 1}^{n} \frac{ w_{\gamma,j} \omega^{\mu}_{\gamma}}{\sigma\sqrt{N}}\Bigg) 
   -\frac{1}{2}  \Bigg(\xi^{\mu}\sum_{\gamma = 1}^{n} 
\frac{ w_{\gamma,j} \omega^{\mu}_{\gamma}}{\sigma\sqrt{N}}\Bigg)^2\Bigg] \Bigg) \right\rangle_{\xi^\mu}\\
		\nonumber &= \prod_{j = 1}^N\left\langle \exp\Bigg(\log \Bigg(1+\ii \Min \Bigg(\xi^{\mu}\sum_{\gamma = 1}^{n} \frac{ w_{\gamma,j} \omega^{\mu}_{\gamma}}{\sigma\sqrt{N}}\Bigg) -\frac{1}{2\sigma^2 N}  \sum_{\beta=1}^{n}w_{\beta,j}\, \omega^{\mu}_{\beta}\sum_{\gamma=1}^{n}w_{\gamma,j}\, \omega^{\mu}_{\gamma} 
  \Bigg)\Bigg)\right\rangle_{\xi^\mu}\\
		\nonumber &=
		\prod_{j = 1}^N \left\langle\exp\Bigg( \ii \Min \Bigg(\xi^{\mu}\sum_{\gamma = 1}^{n} \frac{ w_{\gamma,j} \omega^{\mu}_{\gamma}}{\sigma\sqrt{N}}\Bigg)- \frac{1-\Min^2}{2\sigma^2 N} \sum_{\beta=1}^{n}w_{\beta,j} \omega^{\mu}_{\beta}\sum_{\gamma=1}^{n}{w}_{\gamma,j} \omega^{\mu}_{\gamma}\Bigg)\right\rangle_{\xi^\mu}\\
		&= \left\langle\exp{\Bigg(\ii \Min \xi^{\mu}\sum_{\gamma = 1}^{n} \sum_{j = 1}^{N}\frac{ w_{\gamma,j} \omega^{\mu}_{\gamma}}{\sigma\sqrt{N}}- \frac{1-\Min^2}{2\sigma^2 N}  \sum_{\gamma, \beta=1}^{n} \omega^{\mu}_{\gamma}\omega^{\mu}_{\beta}\sum_{j = 1}^{N}{w}_{\gamma,j} w_{\beta,j} \Bigg)}\right\rangle_{\xi^\mu}\ ,
\end{align}
{the remaining terms vanishing as $O\left(1/\sqrt{N}\right)$.}
{Using $\|\boldsymbol{w}_\gamma\|^2=N$ and setting}
\begin{equation}
\label{params}
M_{\gamma} \equiv \frac{1}{\sqrt{N}}\sum_{j=1}^{N}w_{\gamma,j}\ ,\quad q_{\gamma \beta} \equiv \frac{1}{N}\sum_{j=1}^{N}{w}_{\gamma,j}{w}_{\beta,j} 
\end{equation}
for $\gamma,\beta=1,\dots,n$, $\gamma>\beta$, we finally focus upon		
\begin{equation}
\label{B0mean}
{C(\boldsymbol{\Omega},\boldsymbol{W})}
{\simeq} \prod_{\mu=1}^p\left\langle\exp\Bigg( \frac{\ii \Min\xi^{\mu}}{\sigma}\sum_{\gamma = 1}^{n} M_{\gamma} 
\omega^{\mu}_{\gamma}
-\frac{1-\Min^2}{2\sigma^2} \Bigg[\sum_{\gamma=1}^{n}(\omega_{\gamma}^{\mu})^2   + 2 \sum_{{\gamma> \beta=1}}^{n}\omega_{\beta}^{\mu} \omega_{\gamma}^{\mu} q_{\gamma\beta}\Bigg] \Bigg)
\right\rangle_{\xi^\mu}\ ,
\end{equation}
{neglecting corrections of order $O\left(1/\sqrt{N}\right)$.}
Therefore, {to leading order in $1/N$}, Eq.~\eqref{Bmean} becomes 
	\begin{align}
		\label{ref3}
		 &B(\boldsymbol{Y},\boldsymbol{W}){\simeq}
   \frac{1}{(2\pi)^{nP}}\int_{\mathbb{R}^{2np}}\dd^{nP}\boldsymbol{\Omega}\dd^{nP}\boldsymbol{H}\, \exp\Bigg[-i\sum_{\mu=1}^p\sum_{\gamma = 1}^{n}\left(\eta^\mu_\gamma+\frac{\kappa}{\sigma}\right)\omega_{\gamma}^{\mu}-i\sum_{\mu=1}^p\sum_{\gamma = 1}^{n}y^\mu_\gamma \Phi(\eta^\mu_\gamma)\Bigg]\notag\\
		&\hskip 1cm
  \times \prod_{\mu=1}^p\left\langle\exp\Bigg( \frac{\ii \Min\xi^{\mu}}{\sigma}\sum_{\gamma = 1}^{n} M_{\gamma} 
\omega^{\mu}_{\gamma}
-\frac{1-\Min^2}{2\sigma^2} \Bigg[\sum_{\gamma=1}^{n}(\omega_{\gamma}^{\mu})^2   + 2 \sum_{{\gamma> \beta=1}}^{n}\omega_{\beta}^{\mu} \omega_{\gamma}^{\mu} q_{\gamma\beta}\Bigg] \Bigg)
\right\rangle_{\xi^\mu}\ \Bigg).
  \end{align}
  
Since the integrals for different $\mu$'s are the same and the targets $\xi^\mu$ are statistically independent and equally distributed, Eq.\eqref{Amean} reduces to
  \begin{equation}
  \label{ref3a}
  		A(\boldsymbol{W})=\left(\frac{1}{(2\pi)^{2n}}
    \int_{[1-\epsilon,+\infty)^n}\hskip-1cm\dd^n\boldsymbol{\lambda}\
		\int_{\mathbb{R}^{3n}}\hskip-.5cm\dd^n \boldsymbol{y}\dd^n\boldsymbol{\eta}\dd^n\boldsymbol{\omega}\,
  \left\langle\ee^{{K_\xi\left(\boldsymbol{\eta},\boldsymbol{\lambda},\boldsymbol{y},
  \boldsymbol{\omega}, \boldsymbol{Q},\boldsymbol{M}\right)}}\right\rangle_{\xi}
  \right)^p,
	\end{equation}
	where {$\boldsymbol{\eta}=\{\eta_\gamma\}$, $\boldsymbol{y}=\{y_\gamma\}$, $\boldsymbol{\lambda}
 =\{\lambda_\gamma\}$, $\boldsymbol{\omega}=\{\omega_\gamma\}$, $\boldsymbol{Q}=\{q_{\alpha\beta}\}$, $\boldsymbol{M}=\{M_\gamma\}$,
 for $\alpha,\beta,\gamma=1,\ldots, n$,
 $$
 \dd^n\boldsymbol{\lambda}\equiv \prod_{\gamma=1}^{n}\dd \lambda_\gamma\ ,\ \dd^n\boldsymbol{y}\equiv \prod_{\gamma=1}^{n}\dd y_\gamma\ ,\ \dd^n\boldsymbol{\eta}\equiv \prod_{\gamma=1}^{n}\dd \eta_\gamma\ ,\ 
 \dd^n\boldsymbol{\omega}\equiv \prod_{\gamma=1}^{n}\dd \omega_\gamma
 $$ and} 
	\begin{align}\label{kvar}	
		&K_\xi\left(\boldsymbol{\eta},\boldsymbol{\lambda},\boldsymbol{y},
  \boldsymbol{\omega}, \boldsymbol{Q},\boldsymbol{M}\right)\equiv \ii\sum_{\gamma=1}^{n}\left[y_{\gamma}\lambda_{\gamma}-y_{\gamma} \Phi(\eta_{\gamma}) 
  -\left(\frac{\kappa-\xi \Min M_{\gamma}}{\sigma}+\eta_{\gamma}\right)\omega_{\gamma}\right]\\
\label{kvar1}	
&\hskip 2cm-\frac{1-\Min^2}{2\sigma^2} \Bigg(\sum_{\gamma=1}^{n}(\omega_{\gamma})^2 + 2 \sum_{\gamma\neq \beta=1}^{n}\omega_{\beta} \omega_{\gamma} q_{\gamma\beta}\Bigg).\notag\\
	\end{align}
{Writing the various Dirac deltas appearing in~\eqref{Vmean} as
	\begin{equation}
		\label{delta1}
		\delta \Big(||\boldsymbol{w}_{\gamma}||^2 - N\Big) =  \int_{-\infty}^{+\infty} 
		\frac{\dd E_{\gamma}}{4\pi} \ 
		\ee^{-\ii\frac{E_{\gamma}}{2} N +\ii\frac{E_{\gamma}}{2}\sum_{j=1}^{N}{w}_{\gamma,j}^2},
		\end{equation}
as well as those Dirac deltas whose integration over $q_{\gamma\beta}$, respectively $M_\gamma$, implements the constraints~\eqref{params} within~\eqref{ref3a}, as
		\begin{align}
		\label{delta2}
		\delta \Bigg(q_{\gamma \beta} - \frac{1}{N}\sum_{j=1}^{N}{w}_{\gamma,j}{w}_{\beta,j}\Bigg) &= N \int_{-\infty}^{+\infty} 
		\ \frac{\dd F_{\gamma\beta}}{2\pi} \ 
		\exp\Bigg(-\ii N q_{\gamma \beta}F_{\alpha \beta} + \ii F_{\gamma \beta}\sum_{j=1}^{N}{w_{\gamma,j}w_{\beta,j}}\Bigg)\ ,\\	
	    \label{delta3}
	    \delta\Bigg(M_\gamma - \frac{1}{\sqrt{N}}\sum_{j=1}^{N} {w_{\gamma,j}}\Bigg) 
	    &=\sqrt{N}\int_{-\infty}^{+\infty} 
	    \frac{\dd {L_\gamma}}{2\pi} \ 
	    \exp\Bigg(-\ii\sqrt{N} L_\gamma M_\gamma +\ii\,L_\gamma\sum_{j=1}^{N}w_{\gamma,j}\Bigg) ,
	\end{align}	
and inserting them into~\eqref{Vmean}, one finally arrives at the following explicit integral expression}
	\begin{align*}
&\hskip-.9cm
~\langle {[V^Q_N(\xv^\mu,\xi^\mu,\kappa,\sigma,\epsilon)]}^n \rangle ={\frac{(N)^{n^2/2}}{Z_N^n}\int_{\mathbb{R}^{nN+n^2+2n}} \prod_{\gamma = 1}^{n}\prod_{j = 1}^{N}\prod_{\gamma<\beta=1}^n\dd w_{\gamma, j}\frac{\dd E_{\gamma}}{4\pi}\dd M_\gamma\dd q_{\gamma \beta}\,\frac{\dd F_{\gamma\beta}}{2\pi}\,\frac{\dd {L_\gamma}}{2\pi}}\\ 
&\hskip 1cm\times\exp\Bigg(-i\frac{E_{\gamma}}{2} N +\ii\frac{E_{\gamma}}{2}\sum_{j=1}^{N}w_{\gamma, j}^2\Bigg)\,\exp\Bigg(-\ii N q_{\gamma \beta}F_{\gamma \beta}+\ii F_{\gamma \beta}\sum_{j=1}^{N}w_{\gamma, j}\,w_{\beta, j}\Bigg)\\
&\hskip 2cm
  \times{\exp\Bigg(-\ii\sqrt{N} M_{\gamma}{L_\gamma}+\ii L_\gamma\sum_{j=1}^{N}w_{\gamma, j}\Bigg)}\\
&\hskip 3cm  
\times~{\Bigg[\int_{1-\epsilon}^{+\infty} \frac{1}{(2\pi)^{2n}}
    \int_{[1-\epsilon,+\infty)^n}\hskip-1cm\dd^n\boldsymbol{\lambda}\
		\int_{\mathbb{R}^{3n}}\hskip-.5cm\dd^n \boldsymbol{y}\dd^n\boldsymbol{\eta}\dd^n\boldsymbol{\omega}\,
  \left\langle\ee^{{K_\xi\left(\boldsymbol{\eta},\boldsymbol{\lambda},\boldsymbol{y},
  \boldsymbol{\omega}, \boldsymbol{Q},\boldsymbol{M}\right)}}\right\rangle_{\xi}\Bigg]^p}\ .
  \end{align*}
  {Regrouping together the integrals over $w_{\gamma,j}$ with different $j$ and 
  same $\gamma$, one writes} 
  \begin{align*}
		\nonumber ~\langle {[V^Q_N(\xv^\mu,\xi^\mu,\kappa,\sigma,\epsilon)]}^n \rangle &={\frac{N^{n^2/2}}{Z_N^n}\int_{\mathbb{R}^{n^2+2n}} \prod_{\gamma = 1}^{n}\prod_{\gamma<\beta=1}^n\frac{\dd E_{\gamma}}{4\pi}\dd M_\gamma\dd q_{\gamma \beta}\,\frac{\dd F_{\gamma\beta}}{2\pi}\,\frac{\dd {L_\gamma}}{2\pi}}\\
  &\hskip 2cm
  \times\Bigg[\exp\Bigg({-}\frac{\ii}{2}\sum_{\gamma=1}^{n} E_{\gamma} -\ii\sum_{\gamma<\beta=1}^{n} 
  F_{\gamma \beta}q_{\gamma,\beta}-\frac{\ii}{\sqrt{N}}\sum_{\gamma=1}^n L_\gamma\,M_\gamma\Bigg) \Bigg]^N \\
  &\times\Bigg[\int_{\mathbb{R}^n}  \prod_{\gamma=1}^{n}\dd w_{\gamma} \exp\Bigg(\ii\sum_{\gamma<\beta=1}^{n}F_{\gamma \beta}{w}_{\gamma}{w}_{\beta}  {+}\frac{\ii}{2}\sum_{\gamma=1}^{n}\,E_{\gamma}\,w_{\gamma}^2+\ii \sum_{\gamma=1}^n {L_\gamma}\,w_{\gamma}\Bigg)\Bigg]^N\\
  &\hskip 1cm  
\times~{\Bigg[\frac{1}{(2\pi)^{2n}}
    \int_{[1-\epsilon,+\infty)^n}\hskip-1cm\dd^n\boldsymbol{\lambda}\
		\int_{\mathbb{R}^{3n}}\hskip-.5cm\dd^n \boldsymbol{y}\dd^n\boldsymbol{\eta}\dd^n\boldsymbol{\omega}\,
  \left\langle\ee^{{K_\xi\left(\boldsymbol{\boldsymbol{\eta},\lambda},\boldsymbol{y},
  \boldsymbol{\omega}, \boldsymbol{Q},\boldsymbol{M}\right)}}\right\rangle_{\xi}\Bigg]^p}\ .
	\end{align*}	
	In the large $N$ limit, the contribution $\frac{1}{\sqrt{N}}\sum_{\gamma=1}^n {L_\gamma} M_\gamma$ can be neglected;
 {using~\eqref{ref1}, at leading order in $N\gg1$, one gets}
	\begin{align}
\nonumber		
\langle {[V^Q_N(\xv^\mu,\xi^\mu,\kappa,\sigma,\epsilon)]}^n \rangle&= {\frac{N^{n^2/2}}{Z_N^n}\int_{\mathbb{R}^{3n}}}\left(\prod_{\gamma=1}^{n}\frac{\dd E_{\gamma}}{4\pi}\dd M_\gamma
\frac{\dd L_\gamma}{2\pi}\right) \int_{{\mathbb{R}^{n^2-n}}}~\left(\prod_{\gamma>\beta=1}^n \dd q_{\gamma \beta} \frac{\dd F_{\gamma \beta}}{2\pi }\right) \\
\label{Vaux}  
&\times\exp\Big(N\, G(\mathbf{E},\mathbf{F},\mathbf{L},\mathbf{M},\mathbf{Q})\Big)\ ,
\end{align}
where {$\mathbf{E}=\{E_\gamma\}_{\gamma=1}^n$, $\mathbf{F}=\{F_{\gamma\beta}\}_{\gamma>\beta=1}^n$, 
$\mathbf{L}=\{L_\gamma\}_{\gamma=1}^n$,  and
	\begin{equation}\
 \label{eq:Gdef}
		G(\mathbf{E},\mathbf{F},\mathbf{L},\mathbf{M},\mathbf{Q})\equiv \frac{p}{N}\, 
  G_1(\mathbf{M},\mathbf{Q}) + 
  G_{2}(\mathbf{E},\mathbf{F},\mathbf{L})+
  G_3(\mathbf{E},\mathbf{F},\mathbf{Q})\ ,
	\end{equation}}
with	
	\begin{align}
	\label{G1aux}	&~	{G_1(\mathbf{M},\mathbf{Q})} = \log\left[\int_{{[1-\epsilon,+\infty)^n}} \left(\prod_{\gamma=1}^{n} \frac{\dd \lambda_{\gamma}}{2\pi}\right)
			\int_{~}^{~}\left(\prod_{\gamma=1}^{n}\frac{\dd y_{\gamma}\dd\eta_{\gamma}\dd\omega_{\gamma}}{2\pi}\right)\left\langle\ee^{{K_\xi\left(\boldsymbol{\eta},\boldsymbol{\lambda},\boldsymbol{y},
  \boldsymbol{\omega}, \boldsymbol{Q},\boldsymbol{M}\right)}}\right\rangle_{\xi}\right], \\
\label{G2aux}		&{G_{2}(\mathbf{E},\mathbf{F},\mathbf{L})}= 
  \log\left[\int_{\mathbb{R}^n}^{}  \prod_{\gamma=1}^{n}\dd w_{\gamma} \ee^{\ii\sum_{\gamma<\beta}^{~} F_{\gamma \beta}{w}_{\gamma}{w}_{\beta}  {+}\ii\sum_{\gamma=1}^{n}\frac{E_{\gamma}}{2}\,w_{\gamma}^2\,+\ii\,\sum_{\gamma=1}^{n}w_{\gamma}L_\gamma}\right],\\
\label{G3aux}		& {G_3(\mathbf{E},\mathbf{F},\mathbf{Q})}={-}\frac{\ii}{2}\sum_{\gamma=1}^{n}
   E_{\gamma} -\ii\,\sum_{\gamma>\beta}^{~}F_{\gamma \beta}q_{\gamma,\beta}\ .
	\end{align}
\subsection{Saddle-point approximation}
{When $N$ is large, the behaviour of $\langle [V^Q_N(\xv^\mu,\xi^\mu,\kappa,\sigma,\epsilon)]^n\rangle$ 
can be obtained using the saddle-point approximation, as follows. Setting 
$\mathbf{z}\equiv(\mathbf{E},\mathbf{F},\mathbf{L},\mathbf{M},\mathbf{Q})$ and considering it as a vector in 
$\mathbb{C}^{n^2+2n}$, one expands  $G(\mathbf{z})\equiv G(\mathbf{E},\mathbf{F},\mathbf{L},\mathbf{M},\mathbf{Q})$ around the stationary point $\mathbf{z}^0=(\mathbf{E}^0,\mathbf{F}^0,\mathbf{L}^0,\mathbf{M}^0,\mathbf{Q}^0)$ such that $\displaystyle\frac{\partial G(\mathbf{z}^0)}{\partial z_k}=0$:
$$
G(\mathbf{z})\simeq G(\mathbf{z}^0)+\frac{1}{2}\sum_{j,k=1}^{n^2+2n}\frac{\partial^2G(\mathbf{z}^0)}{\partial z_j
\partial z_k}\,(z_j-z^0_j)\,(z_k-z^0_k)\ .
$$
Let $G''(\mathbf{z}^0)\equiv\left[\frac{\partial^2G(\mathbf{z}^0)}{\partial z_j\partial z_k}\right]$ be the 
Hessian $(n^2+2n)\times (n^2+2n)$ matrix at the stationary point. If such a matrix is negative semi-definite, by suitably deforming the integration paths into the complex domain, one can perform $n^2+2n$ Gaussian integrations by rescaling with $\sqrt{N}$ the corresponding integration variables and approximate:
\begin{equation}\label{eq:aveVnSP}
	\langle[V^Q_N(\xv^\mu,\xi^\mu,\kappa,\sigma,\epsilon)]^n\rangle { \simeq} \frac{1}{Z_N^n}\,\frac{1}{N^n}\,\left(\sqrt{\frac{2\pi}{|\det G''(\mathbf{z}^0)|}}\right)^{n^2+2n}\ \rme^{N G(\mathbf{z}^0)}\ .
\end{equation}
From Eqs. \eqref{eqf7} and \eqref{eq:aveLogV1}, one needs to control the behaviour of the ratio $\displaystyle \frac{1}{nN}\log\langle [(V^Q_N(\xv^\mu,\xi^\mu,\kappa,\sigma,\epsilon)]^n\rangle$ for $n\to0^+$ and $N\to+\infty$. From Eq.~\eqref{eq:aveVnSP} and using Eq.~\eqref{eq:NV} one gets:
\begin{equation}
\label{approxaux}
\frac{1}{nN}\log\langle [V^Q_N(\xv^\mu,\xi^\mu,\kappa,\sigma,\epsilon)]^n\rangle=\frac{1}{n}\,G(\mathbf{z}^0)-\frac{1}{2}\log(2\pi \mathrm{e})+O\left(\frac{\log N}{N}\right)\ .
\end{equation}
\subsection{Replica symmetric ansatz}
Making use of the replica-symmetric \textit{ansatz} which states that the stationary point $\mathbf{z}_0$ is replica-insensitive, one seeks it setting}
	\begin{equation}
		q_{\gamma \beta}=q, \quad F_{\gamma \beta} = F, \quad E_{\gamma}=E\quad   M_{\gamma} = M \quad\text{and}\quad L_\gamma=L  ,
	\end{equation}
	so that~\eqref{kvar} becomes
	\begin{align}
		\nonumber
		{K_\xi\left(\boldsymbol{\eta},\boldsymbol{\lambda},\boldsymbol{y},
  \boldsymbol{\omega}, q,\boldsymbol{M}\right)}&=i\sum_{\gamma=1}^{n}\left[y_{\gamma}\lambda_{\gamma}-y_{\gamma} \Phi(\eta_{\gamma}) 
		-\left(\frac{\kappa-\xi pM}{\sigma}+\eta_{\gamma}\right)\omega_{\gamma}\right]  \\
		\label{kvar2}
		&-\frac{(1-q)(1-\Min^2)}{2\sigma^2} \sum_{\gamma=1 }^{n}\omega_{\gamma}^2-\frac{q(1-\Min^2)}{2\sigma^2}   \Bigg( \sum_{\gamma =1 }^{n}\omega_{\gamma}\Bigg)^2.
	\end{align}
{Notice that the argument of the logarithm in~\eqref{G1aux} is the average of the following quantity
\begin{align}
	{\Delta(n)}&\equiv\int_{{[1-\epsilon,+\infty)^n}} \Bigg(\prod_{\gamma=1}^{n} \frac{\dd \lambda_{\gamma}}{2\pi}\Bigg) \int_{{\mathbb{R}^{2n}}}\Bigg(\prod_{\gamma=1}^{n}\frac{\dd \eta_{\gamma}\dd y_{\gamma}}{2\pi}\Bigg)\int_{\mathbb{R}^n} \prod_{\gamma=1}^{n} \dd \omega_{\gamma} \
  \exp\Bigg(\ii\frac{\xi \Min M}{\sigma}\sum_{\gamma=1}^n\omega_{\gamma}\Bigg)
  \notag  \\
		&\quad\times\exp\Bigg[\ii\sum_{\gamma=1}^{n}\Big(y_{\gamma}\lambda_{\gamma}-y_{\gamma} \Phi(\eta_{\gamma})\Big) -\ii\sum_{\gamma=1}^{n}\Big(\frac{\kappa}{\sigma}+\eta_{\gamma}\Big)\omega_{\gamma}-\frac{(1-q)(1-\Min^2)}{2\sigma^2} \sum_{\gamma=1 }^{n}\omega^2_{\gamma}\notag\\ 
  &\qquad\qquad - \frac{q(1-\Min^2)}{2\sigma^2}\Bigg( \sum_{\gamma =1 }^{n}\omega_{\gamma}\Bigg)^2\Bigg].
  \end{align}
Such a quantity can then be manipulated as follows: using the Gaussian representation
\begin{equation}
\label{gaussrep}
\exp\Bigg(-\frac{q(1-\Min^2)}{2\sigma^2}\Bigg(\sum_{\gamma=1}^n\omega_\gamma\Bigg)^2\Bigg)=\int_{\mathbb{R}}\DD t\exp\Bigg(
-\frac{t^2}{2}-\ii\,t\frac{\sqrt{q(1-\Min^2)}}{\sigma}\sum_{\gamma=1}^n\omega_\gamma\Bigg) \ ,
\end{equation}
via straightforward Gaussian integration over the variables $\omega_\gamma$, one writes}
	\begin{align}
 {\Delta(n)}&=\int_{{[1-\epsilon,+\infty)^2}}\Bigg(\prod_{\gamma=1}^{n} \frac{\dd \lambda_{\gamma}}{2\pi}\Bigg) \int_{{\mathbb{R}^{2n}}}\left(\prod_{\gamma}^{~}\frac{\dd\eta_{\gamma}\dd y_{\gamma}}{2\pi}\right) \prod_{\gamma=1}^{n}\ee^{\ii(  y_{\gamma}\lambda_{\gamma}-y_{\gamma} \Phi(\eta_{\gamma}))} \notag\\
		&\quad\times\int_{{\mathbb{R}}}\DD x \ 
		\prod_{\gamma=1}^{n}\sqrt{\frac{2\pi\sigma^2}{(1-q)(1-\Min^2)}}\,\exp{\left( -\frac{\Big(\kappa-\xi \Min M+\sigma\eta_{\gamma}+x \sqrt{q(1-\Min^2)}\Big)^2 }{2(1-q)(1-\Min^2)} \right)}\ .\notag
	\end{align}
{Notice that the argument of the logarithm in~\eqref{G1aux} amounts to
$\langle\Delta(n)\rangle_\xi$. Then, one sees that $\Delta(n)$ consists in $n$ independent integrals with respect to $\dd\eta_\gamma$, $\dd\lambda_\gamma$ and $\dd y_\gamma$:}	
\begin{align}
		{\Delta(n)}
		&= \int_{{\mathbb{R}}} \DD x \ \Bigg[ \int_{{\mathbb{R}}}\frac{\sigma\,\dd \eta}{\sqrt{2\pi(1-q)(1-\Min^2)}}\int_{1-\epsilon}^{+\infty} \dd \lambda\int_{{\mathbb{R}}}
  \frac{\dd y}{2\pi}\ee^{-\ii y(\lambda-\Phi(\eta/\sigma))}\notag\\
		&\quad\left.\times\exp{\left( -\frac{\Big(\kappa-\xi \Min  M+\sigma\eta+x \sqrt{q(1-\Min^2)}\Big)^2 }
  {2(1-q)(1-\Min^2)} \right)}\right]^n \notag\\
		&= \int_{{\mathbb{R}}} \DD x \ \Bigg[ \int_{{\mathbb{R}}}\frac{\sigma\,\dd \eta}{\sqrt{2\pi(1-q)(1-\Min^2)}}\Theta\left[\Phi\left(\frac{\eta}{\sigma}\right)-1+\epsilon\right] \notag\\
		&\quad\left.\times\exp{\left( -\frac{\Big(\kappa-\xi \Min  M+\sigma\eta+x \sqrt{q(1-\Min^2)}\Big)^2 }{2(1-q)(1-\Min^2)} \right)}\right]^n \notag
\end{align}
{Due to the monotonicity of the error function, the function $\Phi(x)$ in~\eqref{erfunct} is invertible and one has 
$$
\Theta\left[\Phi\left(\frac{\eta}{\sigma}\right)-1+\epsilon\right]=
\Theta\left[\eta-\sigma\Phi^{-1}(1-\epsilon)\right]\ .
$$
Then, by changing the integration variable $\eta$ into $\lambda=\sigma\eta+\kappa$ and using again~\eqref{erfunct}, one gets}

\begin{align}		
		\nonumber
  {\Delta(n)}&=\int_{{\mathbb{R}}}\DD x \ \Bigg[ \int_{\kappa+\sigma\Phi^{-1}(1-\epsilon)}^{+\infty} \frac{\dd \lambda}{\sqrt{2\pi(1-q)(1-\Min^2)}}
		\exp{\left( -\frac{\Big(\lambda-\xi \Min M+x \sqrt{q(1-\Min^2)}\Big)^2 }{2(1-q)(1-\Min^2)} \right)}\Bigg]^n\notag\\
		&=\int_{{\mathbb{R}}} \DD x \ \Bigg[1 - \Phi\Bigg( \frac{\tilde{\kappa}-\xi \Min  M+\sqrt{q(1-\Min^2)}x}{\sqrt{(1-q)(1-\Min^2)}}\Bigg)\Bigg]^n\notag
  \end{align}
  where $\tilde{\kappa}=\kappa+\sigma\Phi^{-1}(1-\epsilon)$. Since the replica trick lets $n$ vanish as a continuous quantity, we can use the first order approximation $z^n\simeq 1+n\log z$, valid for $n\rightarrow 0$, and write:
  \begin{align}
		{\Delta(n)}
		&= 1+n\int_{{\mathbb{R}}} \DD x \ \log{\Bigg[1-\Phi\Bigg( \frac{\tilde{\kappa}-\xi \Min M+\sqrt{q(1-\Min^2)}x}{\sqrt{(1-q)(1-\Min^2)}}\Bigg)\Bigg]}.
	\end{align}

Finally, we notice that the replica symmetric \textit{ansatz} makes all matrix and vector entries equal.
Then, averaging over the target parameter $\xi$ according to the distribution in~\eqref{probpatt} yields the following leading behavior for the function $G_1(M,q)$ when $n\to0^+$:
	\begin{equation}
 \label{eq:G1}{G_1(M,q)}\simeq\log\Big(1+n\,{g(M,q)}\Big)\simeq n\,{g(M,q)}\ ,
 \end{equation} 
where 
\begin{align}
 \nonumber
 &{g(M,q)}\equiv\frac{1+\Mout}{2}\int_{\mathbb{R}} \hskip-.2cm
 \DD x\, \log\left[1-\Phi\left(\frac{x\sqrt{q}-a_-(M)}{\sqrt{1-q}}\right)\right]\\
 &\hskip 2cm
 +\frac{1-\Mout}{2}\int_{\mathbb{R}} \hskip-.2cm
\DD x\,\log\left[1-\Phi\left( \frac{x\sqrt{q}-a_+(M)}{\sqrt{1-q}}\right)\right] \ ,
 \label{gaux}
 \end{align}
 with
 \begin{equation}
 \label{gaux1}
 {a_\pm(M)}\equiv-\,\frac{\tilde{\kappa}\pm\,\Min M}{\sqrt{1-\Min^2}}\ .
 \end{equation}
{Because of the replica symmetric \textit{ansatz}, $G_{2}(E,F,L)$ in~\eqref{G2aux} can be recast as}
\begin{equation}
\label{G2aux1}
{G_{2}(E,F,L)} = \log\left[\int_{\mathbb{R}^n}\dd^n \wv \exp\Bigg(\ii\sum_{\gamma<\beta=1}^n F\,{w}_{\gamma}{w}_{\beta} {+}\ii\,\frac{E}{2}\,\sum_{\gamma=1}^{n}w_{\gamma}^2
+\ii\,L\,\sum_{\gamma=1}^{n}w_{\gamma}\Bigg)\right]\ ,
\end{equation}
where now $\wv=(w_1,\ldots,w_n)$.
Then, rewriting 
$$
\sum_{\gamma<\beta=1}^{n}{w}_{\gamma}{w}_{\beta}=  
\frac{1}{2} \left(\Big(\sum_{\gamma=1}^n {w}_{\gamma}\Big)^2 - \sum_{\gamma=1}^n{w}_{\gamma}^2 \right)\ ,
$$
one finds
$$
\ii\,F\,\sum_{\gamma<\beta=1}^{n}{w}_{\gamma}\,{w}_{\beta}{+}\ii\,\frac{E}{2}\,\sum_{\gamma=1}^{n}w_{\gamma}^2 
=  \ii \frac{F}{2} \Big(\sum_{\gamma=1}^n {w}_{\gamma}\Big)^2 -\ii\,\frac{F{-}E}{2}\sum_{\gamma=1}^n{w}_{\gamma}^2\ ,
$$
and, using~\eqref{gaussrep},
\begin{align}
\exp\Bigg(\ii\sum_{\gamma<\beta=1}^n F\,{w}_{\gamma}{w}_{\beta}  {+}\ii\,\frac{E}{2}\,\sum_{\gamma=1}^{n}w_{\gamma}^2\Bigg) = 
\int_{\mathbb{R}}\frac{\dd x}{\sqrt{2\pi}} \ee^{-x^2/2 -x \sqrt{\ii F}\,\sum_{\gamma}{w}_{\gamma}} \ee^{ -\ii\frac{F{-}E}{2}\sum_{\gamma=1}^{n}w_{\gamma}^2}\ .
	\end{align}
Then, after straightforward Gaussian manipulations and integration,
	\begin{align}
		&{G_{2}(F,E,L)}= \log\left(\int_{\mathbb{R}}\DD x\left(\int_{\mathbb{R}}\dd w\
		\exp\left(-x \sqrt{\ii\, F}\,w-\ii\frac{F{-}E}{2}\,w^2\,+\,\ii\,w\,L\right) \right)^n\right)\notag\\
		&\hskip .5cm 
  =\frac{n}{2}\log\frac{2\pi}{\ii(F{-}E)}\,+\, \log\left( \int_{\mathbb{R}}\DD x \  
  \exp\left(-\ii\,n\,\frac{(x\sqrt{\ii\,F}-\ii\,{L})^2}{2(F{-}E)}\right) \right)\notag\\
		&\hskip 1cm
  =\frac{n}{2}\log\frac{2\pi}{\ii(F{-}E)}\,-\,\frac{1}{2}\log\left(1-n\,\frac{F}{F{-}E}\right) 
  \GG{-}\ii\,\frac{F\,L^2\,n^2}{2(F{-}E)((n-1)F{+}E)}+\ii\,\frac{nL^2}{2(F{-}E)}\ .
  \notag
\end{align}
At leading order in $n\to0^+$ one gets
\begin{align}\label{eq:G2}		
		{G_{2}(F,E,L)} 	&\simeq \frac{n}{2}\left(\log\frac{2\pi}{\ii(F{-}E)}+\frac{F}{F{-}E}\,+\,\ii\frac{L^2}{F{-}E}\right)\ .
\end{align}
Finally, the replica symmetry \textit{ansatz}  yields the following leading order behaviour for~\eqref{G3aux},
	\begin{align}\label{eq:G3}
	{G_3(E,F,q)}\simeq {-}\ii\,n \frac{E}{2} - \ii\,F\,q\,\frac{n(n-1)}{2} 
  \simeq {-}\ii\,n\,\frac{1}{2}(E {-} Fq) \ .
	\end{align}

Using \eqref{eq:G1},~\eqref{eq:G2} and \eqref{eq:G3} into~\eqref{eq:Gdef}, we get, at the leading order in $n\rightarrow 0^+$:
\begin{align}
 \label{Gauxfinal}
	&{G(E,F,L,M,q)} \simeq\,n\,{\alpha\,g(M,q)}
 \,+\,\frac{n}{2}\left( \log\frac{2\pi}{\ii(F{-}E)}\,+\,\frac{F}{F{-}E}+\,\ii\frac{L^2}{F{-}E}\,{+}\ii(q\,F{-}E)\right)\ .
	\end{align}
The stationary point $\mathbf{z}^0$ is then found by asking that 
$\frac{\partial G(\mathbf{z})}{\partial E} =\frac{\partial G(\mathbf{z})}{\partial F} =\frac{\partial G(\mathbf{z})}{\partial L} = 0$ yielding the stationary point components:
	\begin{equation}
		F^0= -\frac{\ii q}{(1-q)^2}\ , \quad {E^0 = \ii\frac{1-2q}{(1-q)^2}}\ ,\quad L^0 =0\ .
	\end{equation}
	
	Then, from \eqref{Gauxfinal} one obtains
	\begin{equation}
		\label{eq2}
		{\frac{G(E^0,F^0,L^0,M,q)}{n}}=\alpha\,{g(M,q)}
  +\frac{1}{2}\Big(\log(2\pi(1-q))\,+\,\frac{1}{(1-q)}\Big)\ .
	\end{equation}
{The sought after stationary point of $G(\mathbf{z})$ is finally obtained by setting
\begin{equation}
\label{finalstationary}
\partial_q {G(E^0,F^0,L^0,M,q)}=\partial_M {G(E^0,F^0,L^0,M,q)}=0\ .
\end{equation}
Using~\eqref{gaux}, one explicitly computes, with $\beta=q,M$, 
\begin{align}
\nonumber
&\partial_\beta\, g(M,q)=-\frac{1+\Mout}{2\sqrt{2\pi}}\int_{\mathbb{R}}\DD x\, 
\frac{\exp\left(-\frac{(x\sqrt{q}-a_-(M))^2}{2(1-q)}\right)}{1-\Phi\left(\frac{x\sqrt{q}-a_-(M)}{\sqrt{1-q}}\right)}
\,\partial_\beta\frac{x\sqrt{q}-a_-(M)}{\sqrt{1-q}}
\\
\label{stableaux}
&\hskip 4cm
-\frac{1-\Mout}{2\sqrt{2\pi}}\int_{\mathbb{R}}\DD x\, \frac{\exp\left(-\frac{(x\sqrt{q}-a_+(M))^2}{2(1-q)}\right)}{1-\Phi\left(\frac{x\sqrt{q}-a_+(M)}{\sqrt{1-q}}\right)}\,\partial_\beta\frac{x\sqrt{q}-a_+(M)}{\sqrt{1-q}}\ .
\end{align}
As observed at the end of Section~\ref{sec:storcap}, the optimal storage capacity is obtained when $q$ in~\eqref{params}, namely the average overlap of random weights, tends to $1$. The condition $q\rightarrow 1$ implies that the arguments of the functions $\Phi$ in~\eqref{gaux} tend to $\pm\infty$. Then, one can use the asymptotic behavior of the error function,
\begin{equation}
\label{asymperr}
\hbox{erf}(x)\simeq\pm 1-\frac{{\rm e}^{-x^2}}{\sqrt{\pi}}\left(\frac{1}{x}-\frac{1}{2x^3}\right)\ \hbox{when}\quad x\to\pm\infty\ ,
\end{equation}
together with~\eqref{erfunct} to get that
\begin{equation}
\label{asympPHI}
\frac{1}{1-\Phi\left(\frac{x\sqrt{q}-a_\pm(M)}{\sqrt{1-q}}\right)}\simeq\left\{\begin{matrix}\exp\left(\frac{(x\sqrt{q}-a_\pm(M))^2}{2(1-q)}\right)\,\sqrt{2\pi}\frac{x\sqrt{q}-a_\pm(M)}{\sqrt{1-q}}&\ldots&x\sqrt{q}-a_\pm(M)> 0\cr
1&\ldots& x\sqrt{q}-a_\pm(M)\leq 0
\end{matrix}\right.\ .
\end{equation}
Therefore, for $q\to 1^-$, the vanishing Gaussian terms in~\eqref{stableaux} can only be compensated for $x\sqrt{q}\geq a_\pm(M)$, so that
\begin{align}
\nonumber
&\partial_\beta\, g(M,q)\simeq-\frac{1+\Mout}{2}\int_{a_-(M)/\sqrt{q}}^{+\infty}\DD x\, \frac{x\sqrt{q}-a_-(M)}{\sqrt{1-q}}
\,\partial_\beta\frac{x\sqrt{q}-a_-(M)}{\sqrt{1-q}}
\\
\label{stableaux2}
&\hskip 4cm
-\frac{1-\Mout}{2}\int_{{a_+(M)/\sqrt{q}}}^{{+\infty}}\DD x\, \frac{x\sqrt{q}-a_+(M)}{\sqrt{1-q}}\,
\partial_\beta\frac{x\sqrt{q}-a_+(M)}{\sqrt{1-q}}\ .
\end{align}
Furthermore, using~\eqref{gaux1},
\begin{align}
\nonumber
&\partial_M\, g(M,q)\simeq\frac{\Min}{\sqrt{(1-\Min^2)(1-q)}}\Bigg(-\frac{1+\Mout}{2}\int_{a_-(M)/\sqrt{q}}^{+\infty}\DD x\,(x\sqrt{q}-a_-(M))
\\
\label{stableaux3}
&\hskip 4cm
+\frac{1-\Mout}{2}\int_{a_+(M)/\sqrt{q}}^{{+\infty}}\DD x\, (x\sqrt{q}-a_+(M))\Bigg)\ ,
\end{align}
which, together with~\eqref{finalstationary} yields \eqref{eq:alpha2}.
On the other hand,~\eqref{eq2} and~\eqref{finalstationary} yields
\begin{equation}
\label{stableaux4}
\alpha\,\partial_q\,g(M,q)=-\frac{q}{2(1-q)^2}\ .
\end{equation}
Thus, from 
$$
\partial_q\frac{x\sqrt{q}-a_\pm(M)}{\sqrt{1-q}}=\frac{x}{2\sqrt{q(1-q)}}-\frac{x\sqrt{q}-a_\pm(M)}{2(1-q)^{3/2}}\ ,
$$
and~\eqref{stableaux2} one retrieves~\eqref{eq:alpha1}} {as the leading term in $(1-q)^{-1}$ in the limit $q\rightarrow 1^{-}$}.

\section{Large output bias limit}
\label{appB}

{ 
In order to extract the leading order behaviour of the quantum critical storage capacity when the target bias
$\Mout\to1^-$, we distinguish two possibilities.
Firstly, in this Appendix, we keep the input bias $\Min$ fixed and let $\Mout\to1^-$; then, in the next one we treat the case 
when $\Min=\Mout=m\to 1^-$. 
Consider the equation 
$$
(1+\Mout) \int_{a_-(M)}^{+\infty} \DD x\,(x-a_-(M))=(1-\Mout) \int_{a_+(M)}^{+\infty}\DD x \, (x-a_+(M))\ ,
$$
with $a_\pm(M)$ as in~\eqref{apm}.
If $\Mout\to1^-$ and $\Min$ is kept fixed, $M$ must diverge to $+\infty$. Otherwise, the left-hand side 
cannot vanish, being the integral of a positive function. Then, in the following, we shall consider 
$\Mout$ close to $1$ so that $a_+(M)<0$, $a_-(M)>0$ and
$$
\mp a_\pm(M)=\frac{\Min\,M}{\sqrt{1-\Min^2}}\pm\frac{\tilde{\kappa}}{\sqrt{1-\Min^2}}\gg 1\ ,
$$
yielding 
\begin{equation}
\label{lowboundinM}
M\gg\frac{\sqrt{1-\Min^2}}{\Min}\,+\,\frac{\tilde{\kappa}}{\Min}.
\end{equation}
In the classical case $\tilde{\kappa}=\kappa$; moreover, the limit behaviour of the critical 
classical storage capacity for $\Mout\to1^+$,
$$
\alpha^C_c\simeq-\frac{1}{(1-\Mout)\,\log(1-\Mout)}\ ,
$$
is obtained for $\kappa=0$, namely for $\displaystyle M\gg\frac{\sqrt{1-\Min^2}}{\Min}$.}

We proceed by recasting the two storage capacity defining equations~\eqref{eq:alpha1} and~\eqref{eq:alpha2} in terms of Gaussian and error functions:
\begin{align}
\nonumber
&\alpha_c^Q\Bigg[ \frac{1+\Mout}{4} \Bigg((1+a_-^2(M))\,\left(1-{\rm erf}\left(\frac{a_-(M)}{\sqrt{2}}\right)\right)-
a_-(M)\,\sqrt{\frac{2}{\pi}}{\rm e}^{-a_-^2(M)/2}\Bigg)\\
&\hskip 2cm
\label{eq:alpha1aux}
+\frac{1-\Mout}{4}\Bigg((1+a_+^2(M))\,\left(1-{\rm erf}\left(\frac{a_+(M)}{\sqrt{2}}\right)\right)
-a_+(M)\,\sqrt{\frac{2}{\pi}}{\rm e}^{-a_+^2(M)/2}\Bigg)\Bigg] = 1\ ,
\end{align}
respectively
\begin{equation}
\label{eq:alpha2aux}
		\frac{1-\Mout}{1+\Mout}=\frac{\sqrt{\frac{2}{\pi}}\,{\rm e}^{-a_-^2(M)/2}-a_-(M)\left(1-\hbox{erf}\left(\frac{a_-(M)}{\sqrt{2}}\right)\right)}{\sqrt{\frac{2}{\pi}}\,{\rm e}^{-a_+^2(M)/2}-a_+(M)\left(1-\hbox{erf}\left(\frac{a_+(M)}{\sqrt{2}}\right)\right)}\ .
\end{equation}
Equation \eqref{eq:alpha2aux} can be satisfied in the limit $\Mout\rightarrow 1^{-}$ only if the right-hand side vanishes, which can happen only for values of $M$ such that $a_\pm(M)\to \mp \infty$. For small but finite values of $1-\Mout$, the solution of \eqref{eq:alpha2aux} is obtained for $\mp a_\pm(M)\gg1$, which implies
\begin{equation}
M\gg\frac{\sqrt{1-\Min^2}}{\Min}\,+\,\frac{\tilde{\kappa}}{\Min}.
\end{equation}
Using the asymptotic behaviour in~\eqref{asymperr} with $\mp a_\pm(M)\gg1$, one gets
\begin{align}
\label{asymp1}
1-\hbox{erf}\left(\frac{a_-(M)}{\sqrt{2}}\right)\simeq\sqrt{\frac{2}{\pi}}{\rm e}^{-a_-^2(M)/2}\,\left(
\frac{1}{a_-(M)}-\frac{1}{a_-^3(M)}\right)\ ,
\\
\label{asymp2}
1-\hbox{erf}\left(\frac{a_+(M)}{\sqrt{2}}\right)\simeq 2+\sqrt{\frac{2}{\pi}}{\rm e}^{-a_+^2(M)/2}\,\left(
\frac{1}{a_+(M)}-\frac{1}{a_+^3(M)}\right){\simeq 2}\ .
\end{align}
Hence, for $\Mout\to1^-$, at leading order,~\eqref{eq:alpha2aux} and~\eqref{eq:alpha1aux} read
\begin{equation}
\label{asymp3}
{1-\Mout\simeq -\frac{{\rm e}^{-a_-^2(M)/2}}{\sqrt{\frac{\pi}{2}}a_-^2(M) a_+(M)}}\ ,\quad 
{1\simeq\alpha_c^Q\,\frac{1-\Mout}{2}\,a_+^2(M)}\ .
\end{equation}
Since $\Mout\to1^-$ implies $a_-(M)\simeq -\,a_+(M)$, the first asymptotic behaviour in~\eqref{asymp3} yields 
\begin{equation}
\label{asymp5}
{\log(1-\Mout)\simeq-\frac{a_-^2(M)}{2}-\log\left(\sqrt{\frac{\pi}{2}}a_-^2(M)(-a_+(M))\right)\simeq
-\,\frac{a_-^2(M)}{2}\simeq -\,\frac{a_+^2(M)}{2}}\ .
\end{equation}
from which  the limit of strongly correlated targets, $\Mout\to1^-$, is as in~\eqref{asymp0}, for all pattern biases $0\leq \Min\leq 1$.
{However, it must be emphasized that, because of~\eqref{lowboundinM} the limit is reached with possibly quite different slopes depending on both the quantum parameter $\sigma\Phi^{-1}(1-\epsilon)$ and on the degree of independence of the input patterns $\Min$.}

\section{Simultaneously large input and output bias}
\label{appC}

 Setting $\Min=\Mout=m$, the quantity $M$ has to be chosen such that
 \begin{equation}
\label{appC1}
		\frac{1-m}{1+m}=\frac{\sqrt{\frac{2}{\pi}}\,{\rm e}^{-a_-^2(M)/2}-a_-(M)\left(1-\hbox{erf}\left(\frac{a_-(M)}{\sqrt{2}}\right)\right)}{\sqrt{\frac{2}{\pi}}\,{\rm e}^{-a_+^2(M)/2}-a_+(M)\left(1-\hbox{erf}\left(\frac{a_+(M)}{\sqrt{2}}\right)\right)}=\frac{I_-(M)}{I_+(M)}\ ,
\end{equation}
where we set
\begin{equation}
\label{appC2}
 I_\pm(M)=\sqrt{\frac{2}{\pi}}\,{\rm e}^{-a_\pm^2(M)/2}-a_\pm(M)\left(1-\hbox{erf}\left(\frac{a_\pm(M)}
{\sqrt{2}}\right)\right)   
\end{equation}
In the limit $m\to1^-$, the quantities $a_\pm(M)$ in~\eqref{apm} behave as
 \begin{equation}
 \label{appC3}
 a_\pm(M)\simeq-\frac{\tilde{\kappa}\pm M}{\sqrt{2(1-m)}}
 \end{equation}
 and both diverge unless $M=\tilde{\kappa}$. Notice however, that inserting  $M=\tilde{\kappa}$ into~\eqref{appC1} and taking the limit, the equality~\eqref{appC1} cannot be satisfied. Indeed, 
 $a_-(\tilde{\kappa})=0$ and
 $$
 a_+(\tilde{\kappa})\simeq-\frac{2\tilde{\kappa}}{\sqrt{2(1-m)}}\Longrightarrow
 \frac{I_-(M)}{I_+(M)}\simeq\sqrt{1-m}\ ,
 $$
 whereas the right-hand side vanishes as $1-m$. Therefore, we need first to consider the asymptotic behaviour of the right-hand side of~\eqref{appC1} when $m\to1^-$ and then properly choose $M$ which will thus depend on $m$.
 
 We distinguish the following cases
 \begin{align}
\label{case1}
M>\tilde{\kappa}&\Longrightarrow\left\{\begin{matrix}a_-(M)>0&\Rightarrow& I_-(M)&\simeq&\sqrt{\frac{2}{\pi}}{\rm e}^{-a^2_-(M)/2}\frac{1}{a_-^2(M)}\cr
a_+(M)<0&\Rightarrow& I_+(M)&\simeq&-2a_+(M)
\end{matrix}
\right.\ ,\\
\label{case2}
 -\tilde{\kappa}<M<\tilde{\kappa}&\Longrightarrow\left\{\begin{matrix}a_-(M)<0&\Rightarrow& I_-(M)&\simeq&-2a_-(M)
 \cr
a_+(M)<0&\Rightarrow& I_+(M)&\simeq&-2a_+(M)
\end{matrix}
\right.\ ,\\
\label{case3}
 M<-\tilde{\kappa}&\Longrightarrow\left\{\begin{matrix}a_-(M)<0&\Rightarrow& I_-(M)&\simeq&-2a_-(M)\cr
a_+(M)>0&\Rightarrow& I_+(M)&\simeq&\sqrt{\frac{2}{\pi}}{\rm e}^{-a^2_+(M)/2}\frac{1}{a_+^2(M)}
\end{matrix}
\right.\ ,
\end{align}
where we made explicit the asymptotic behaviours~\eqref{asymp1} and~\eqref{asymp2} when $m\to1^-$.
Then, one finds
\begin{align}
\label{case1a}
M>\tilde{\kappa}&\Longrightarrow&\frac{I_-(M)}{I_+(M)}&\simeq&\frac{2}{\sqrt{\pi}}
\frac{(1-m)^{3/2}}{(\tilde{\kappa}-M)^2(\tilde{\kappa}+M)}\exp\left(-\frac{(\tilde{\kappa}-M)^2}{4(1-m)}\right)\ ,\\
\label{case2a}
 -\tilde{\kappa}<M<\tilde{\kappa}&\Longrightarrow&\frac{I_-(M)}{I_+(M)}&\simeq&\frac{\tilde{\kappa}-M}{\tilde{\kappa}+M}
\ ,\\
\label{case3a} 
M<-\tilde{\kappa}&\Longrightarrow&\frac{I_-(M)}{I_+(M)}&\simeq&
\frac{\sqrt{\pi}}{2}
\frac{(\tilde{\kappa}-M)^2(\tilde{\kappa}+M)}{(1-m)^{3/2}}
\exp\left(+\frac{(\tilde{\kappa}-M)^2}{4(1-m)}\right)\ .
\end{align}
Since~\eqref{appC1} asks for $\displaystyle\frac{I_-(M)}{I_+(M)}\simeq\frac{1-m}{2}$, together with $a_-(M)\to+\infty$ and 
$a_+(M)\to-\infty$, the only behaviour compatible with these request is the one in~\eqref{case1a}.
Indeed, the third one is clearly to be excluded, while the second one 
asks for
\begin{equation}
\label{appC4}
M\simeq\tilde{\kappa}\,m\Longrightarrow\left\{
\begin{matrix}
a_-(M)&\simeq&-\tilde{\kappa}\sqrt{\frac{1-m}{2}}\cr
a_+(M)&\simeq&-\tilde{\kappa}\sqrt{\frac{2}{1-m}}
\end{matrix}\right.\ ,
\end{equation}
which requires $a_-(M)\to0$ instead of $a_-(M)\to+\infty$.
Then, the only possible remaining behaviour yields
\begin{equation}
\label{appC4a}
    \frac{4}{\sqrt{\pi}}
\frac{(1-m)^{1/2}}{(\tilde{\kappa}-M)^2(\tilde{\kappa}+M)}\exp\left(-\frac{(\tilde{\kappa}-M)^2}{4(1-m)}\right)\simeq 1
\end{equation}
or, in terms of $a_\pm(M)$,
\begin{equation}
\label{appC4b}
    \sqrt{\frac{2}{\pi}}
\frac{1}{a_-(M)}\exp\left(-\frac{a_{-}^2(M)}{2}\right)\simeq a_-(M) a_+(M) (1-m)\ .
\end{equation}
The functional dependence of $M$ on $m$ when $m\to1^-$, implicitly determined by~\eqref{appC4a}, cannot be given in terms of simple functions and can be obtained only numerically; however,~\eqref{appC4a} implies that
\begin{equation}
\label{appC4c}
\lim_{m\to 1^-}M(m)=\widetilde{\kappa}\ .
\end{equation}
Finally, using~\ref{appC4b} and~\eqref{appC4c},~\eqref{asymp1} and~\ref{asymp2}, together with~\eqref{appC3}, from~\eqref{eq:alpha1aux}, one gets
\begin{eqnarray}
\nonumber
\frac{1}{\alpha^Q_c}&\simeq&\Bigg[-\frac{1}{\sqrt{2\pi}}\frac{1}{a_-^3(M)}\exp\left(-\frac{a_-^2(M)}{2}\right)
+\frac{1-m}{2}\,a_+^2(M)\Bigg]=\frac{1-m}{2}\frac{a_+(M)}{a_-(M)}\Big(a_-(M)a_+(M)-1\Big)\\
\label{appC4d}
&\simeq&\frac{1-m}{2}\,\frac{\widetilde{\kappa}+M}{\widetilde{\kappa}-M}\,\frac{\widetilde{\kappa}^2-M^2}{2(1-m)}\,=\,\frac{\Big(\widetilde{\kappa}+M\Big)^2}{4}\simeq\widetilde{\kappa}^2\ .
\end{eqnarray}

\end{document}